\documentclass[aps,prb,preprint,showpacs,floatfix,superscriptaddress]{revtex4}

\usepackage{epsfig,amsmath,amssymb} 
\usepackage{graphics}
\usepackage{multirow}
\usepackage{epsfig}
\usepackage{subfig}
\usepackage{psfig}
\usepackage{color}

\usepackage{multirow}

\begin{document}

\title
{
Magnetic order on a frustrated spin-$\frac{1}{2}$ Heisenberg antiferromagnet on the Union Jack lattice
}
\author
{
R.~F.~Bishop 
}
\affiliation
{School of Physics and Astronomy, Schuster Building, The University of Manchester, Manchester, M13 9PL, UK}

\author
{
P.~H.~Y.~Li
}
\affiliation
{School of Physics and Astronomy, Schuster Building, The University of Manchester, Manchester, M13 9PL, UK}

\author 
{
D.~J.~J.~Farnell
} 
\affiliation {Health Methodology Research Group, School of Community-Based Medicine, Jean McFarlane Building, University Place, The University of Manchester, M13 9PL, UK}

\author
{
C.~E.~Campbell
}
\affiliation
{School of Physics and Astronomy, University of Minnesota, 116 Church Street SE, Minneapolis, Minnesota 55455, USA}

\begin{abstract}
  We use the coupled cluster method (CCM) to study the
  zero-temperature phase diagram of a two-dimensional frustrated
  spin-half antiferromagnet, the so-called Union Jack model. It is
  defined on a square lattice such that all nearest-neighbor pairs are
  connected by bonds with a strength $J_{1} > 0$, but only half the
  next-nearest-neighbor pairs are connected by bonds with a strength
  $J_{2} \equiv \kappa J_{1} > 0$. The bonds are arranged such that on
  the $2 \times 2$ unit cell they form the pattern of the Union Jack
  flag. Alternating sites on the square lattice are thus 4-connected
  and 8-connected. We find strong evidence for a first phase
  transition between a N\'{e}el antiferromagnetic phase and a canted
  ferrimagnetic phase at a critical coupling $\kappa_{c_{1}} = 0.66
  \pm 0.02$. The transition is an interesting one, at which the energy
  and its first derivative seem continuous, thus providing a typical
  scenario of a second-order transition (just as in the classical case
  for the model), although a weakly first-order transition cannot be
  excluded. By contrast, the average on-site magnetization approaches
  a nonzero value $M_{c_{1}}=0.195 \pm 0.005$ on both sides of the
  transition, which is more typical of a first-order transition. The
  slope, $dM/d\kappa$, of the order parameter curve as a function of
  the coupling strength $\kappa$, also appears to be continuous, or
  very nearly so, at the critical point $\kappa_{c_{1}}$, thereby
  providing further evidence of the subtle nature of the transition
  between the N\'{e}el and canted phases. Our CCM calculations provide
  strong evidence that the canted ferrimagnetic phase becomes unstable
  at large values of $\kappa$, and hence we have also used the CCM
  with a model collinear semi-stripe-ordered ferrimagnetic state in
  which alternating rows (and columns) are ferromagnetically and
  antiferromagnetically ordered, and in which the spins connected by
  $J_{2}$-bonds are antiparallel to one another. We find tentative
  evidence, based on the relative energies of the two states, for a
  second zero-temperature phase transition between the canted and
  semi-stripe-ordered ferrimagnetic states at a large value of the
  coupling parameter around $\kappa_{c_{2}} \approx 125 \pm 5$. This
  prediction, however, is based on an extrapolation of the CCM results
  for the canted state into regimes where the solutions have already
  become unstable and the CCM equations based on the canted state at
  any level of approximation beyond the lowest have no solutions. Our
  prediction for $\kappa_{c_{2}}$ is hence less reliable than that for
  $\kappa_{c_{1}}$. Nevertheless, if this second transition at
  $\kappa_{c_{2}}$ does exist, our results clearly indicate it to be
  of first-order type.
\end{abstract}

\pacs{75.10.Jm, 75.30.Gw, 75.40.-s, 75.50.Ee}

\maketitle

\section{Introduction}
\label{Introd}
Quantum magnetism at zero temperature for lattices in two spatial
dimensions \cite{2D_magnetism_1,2D_magnetism_2,2D_magnetism_3} is an
important and fascinating subject because such systems display a wide
variety of behavior, including semi-classical N\'eel ordering,
two-dimensional quantum ``spirals'', valence-bond crystals/solids, and
spin liquids. The behavior of these systems is driven by the 
nature of the underlying crystallographic lattice, the number and
range of bonds on this lattice, and the spin quantum numbers of the
atoms localised to the sites on the lattice. There are very few exact
results for quantum spin systems on two-dimensional (2D) lattices, and
so the application of approximate methods is crucial to their
understanding. The theoretical investigation of these models has been
strongly mirrored by the discovery and experimental investigation of
new quasi-2D magnetic materials. It seems clear that we can only form
a complete picture of such 2D quantum spin-lattice systems by
considering a wide range of possible scenarios that are often inspired
(or followed shortly afterwards) by experimental studies.

A prototypical case is presented by the spin-half square-lattice
Heisenberg antiferromagnet (HAF) model. This model has been studied
extensively via a range of approximate techniques.
\cite{Sa:1997,square_swt,square_series,2D_magnetism_3} Its basic
properties have been well-established, where, for example, approximate
results for the order parameter indicate that about 61\% of the
classical N\'{e}el ordering persists in the quantum limit at zero
temperature. A review of the properties of the spin-half
square-lattice HAF is given in
Ref. [\onlinecite{square_manousakis}]. The most accurate results for
this model are provided by quantum Monte Carlo (QMC)
simulations.\cite{Sa:1997} Indeed, QMC techniques generally provide
the benchmark for quantum magnets in two spatial dimensions. However,
its use is severely limited by the ``sign problem,'' which is often a
consequence of quantum frustration in the context of lattice spin
systems.

A common theme has also begun to emerge recently when frustrating
next-nearest-neighbor (NNN) bonds with strength $J_{2} > 0$ are added
to the basic spin-half square-lattice HAF with nearest-neighbor (NN)
bonds with strength $J_{1} > 0$. The frustrating $J_{2}$ bonds may be
added on some or all of the square plaquettes of the lattice and/or
across both or only one of the diagonals of each plaquette. Perhaps
the prototypical such model is the so-called spin-half
$J_{1}$--$J_{2}$ model in which all possible NNN bonds are
included. Recent interest in this model has been reinvigorated by the
discovery of various layered magnetic materials, such as
Li$_2$VOSiO$_4$, Li$_2$VOGeO$_4$, VOMoO$_4$, and
BaCdVO(PO$_4$)$_2$. Several approximate methods have been used to
simulate the properties of this system including the coupled cluster
method (CCM),
\cite{j1j2_square_ccm1,Bi:2008_PRB,Bi:2008_JPCM,j1j2_square_ccm4,j1j2_square_ccm5}
series expansion (SE) techniques,
\cite{j1j2_square_series1,j1j2_square_series2,j1j2_square_series3,j1j2_square_series4,j1j2_square_series5}
exact diagonalization (ED) methods,
\cite{j1j2_square_ed1,j1j2_square_ed2,j1j2_square_ed3} and
hierarchical mean-field (MF) calculations.\cite{j1j2_square_mf} These
approximate techniques have established conclusively that there are
two phases exhibiting magnetic long-range order (LRO) at small and at
large values of $\kappa \equiv J_{2}/J_{1}$ respectively.  For $\kappa
< \kappa_{c_{1}} \approx 0.4$ the ground-state (gs) phase exhibits
(NN) N\'{e}el magnetic LRO, whereas for $\kappa > \kappa_{c_{2}}
\approx 0.6$ it exhibits collinear striped LRO in which alternating
rows (or columns) of the square lattice have opposite spins, with the
spins on each row (or columns) aligned, so that the N\'{e}el order is
between NNN pairs. The intermediate region consists of a quantum
paramagnetic state without magnetic LRO.

Several other models in this general class of spin-half models with
both NN and NNN interactions have prompted recent interest. They all
involve the removal of some of the NNN $J_{2}$-bonds from the
fundamental $J_{1}$--$J_{2}$ model. One such example is the
Shastry-Sutherland model,\cite{shastry1,shastry2,shastry3} realized
experimentally by the magnetic
material $\mathrm{SrCu}(\mathrm{BO}_3)_2$, which involves the removal
of three-quarters of the $J_{2}$-bonds. Whereas the $J_{1}$--$J_{2}$
model on the 2D square lattice has each of the sites connected by 8
bonds (4 NN $J_{1}$-bonds and 4 NNN $J_{2}$-bonds) to other sites, the
Shastry-Sutherland model has each of the sites connected by 5 bonds (4
NN $J_{1}$-bonds and 1 NNN $J_{2}$-bond). A second example is
the HAF on the anisotropic triangular lattice model (also known as the
interpolating square-triangle model),\cite{square_triangle} realized
experimentally by the magnetic material Cs$_2$CuCl$_4$, which involves
the removal of half the $J_{2}$-bonds from the original
$J_{1}$--$J_{2}$ model. In this model each of the sites on the 2D
square lattice is connected by 6 bonds (4 NN $J_{1}$-bonds and 2 NNN
$J_{2}$-bonds) to other sites, such that the remaining $J_{2}$-bonds
connect equivalent NNN sites in each square plaquette. Although
all of the models mentioned above show antiferromagnetic N\'eel
ordering for small $J_2$, their phase diagrams for larger $J_2$
display a wide variety of behavior, including, two-dimensional quantum
``spirals'', valence-bond crystals/solids, and spin liquids. Thus, in
the absence of any definitive theoretical argument, the best way to
understand this class of NN/NNN models on the square lattice is to
treat each one on a case-by-case basis.

In this article we study another frustrated spin-half model that has
both NN ($J_1$) and NNN ($J_2$) bonds on the square lattice, where
these bonds form a pattern that resembles the ``Union Jack''
flag. Just as for the anisotropic triangular HAF described above, the
Union Jack model on the 2D square lattice also has only one
frustrating NNN bond per square plaquette, but these $J_{2}$-bonds are
now arranged such that half the sites are connected by 8 bonds (4 NN
$J_{1}$-bonds and 4 NNN $J_{2}$-bonds) to other sites, while the other
half are connected only by 4 $J_{1}$-bonds to their NN sites, as
described more fully below in Sec.\ \ref{model_section}. This model
has previously been studied using SWT\cite{Co:2006_PRB,Co:2006_JPhys}
and SE techniques.\cite{Zh:2007} As in the case of the spin-half
interpolating square-triangle model, it was
shown\cite{Co:2006_PRB,Co:2006_JPhys,Zh:2007} that NN N\'eel order for
the Union Jack model persists until a critical value of the
frustrating NNN ($J_2$) bonds. However, in contrast to the
interpolating square-triangle model, there exists a ferrimagnetic
ground state in which spins on the eight-connected sites cant at a
nonzero angle with respect to their directions in the corresponding
N\'{e}el state. This model thus exhibits an overall magnetic moment in
this regime, which is quite unusual for spin-half 2D materials with
only Heisenberg bonds and which therefore preserve (spin) rotational
symmetries in the Hamiltonian. This model also presents us with a
difficult computational task in order to simulate its properties.
Here we wish to study this model using the CCM, which has consistently
been shown to yield insight into a wide range of problems in quantum
magnetism, and which we now hope will hence shed yet more light on the
whole class of NN/NNN models as mentioned before. The dual associated
features of a model with two sorts of sites with differing
connectivities, and its consequent ferrimagnetic phase, are just those
that are attracting the interest of the community now.

\section{The model}
\label{model_section}
\begin{figure}[t]
\mbox{
   \subfloat[Canted\label{canted_model}]{\scalebox{0.45}{\epsfig{file=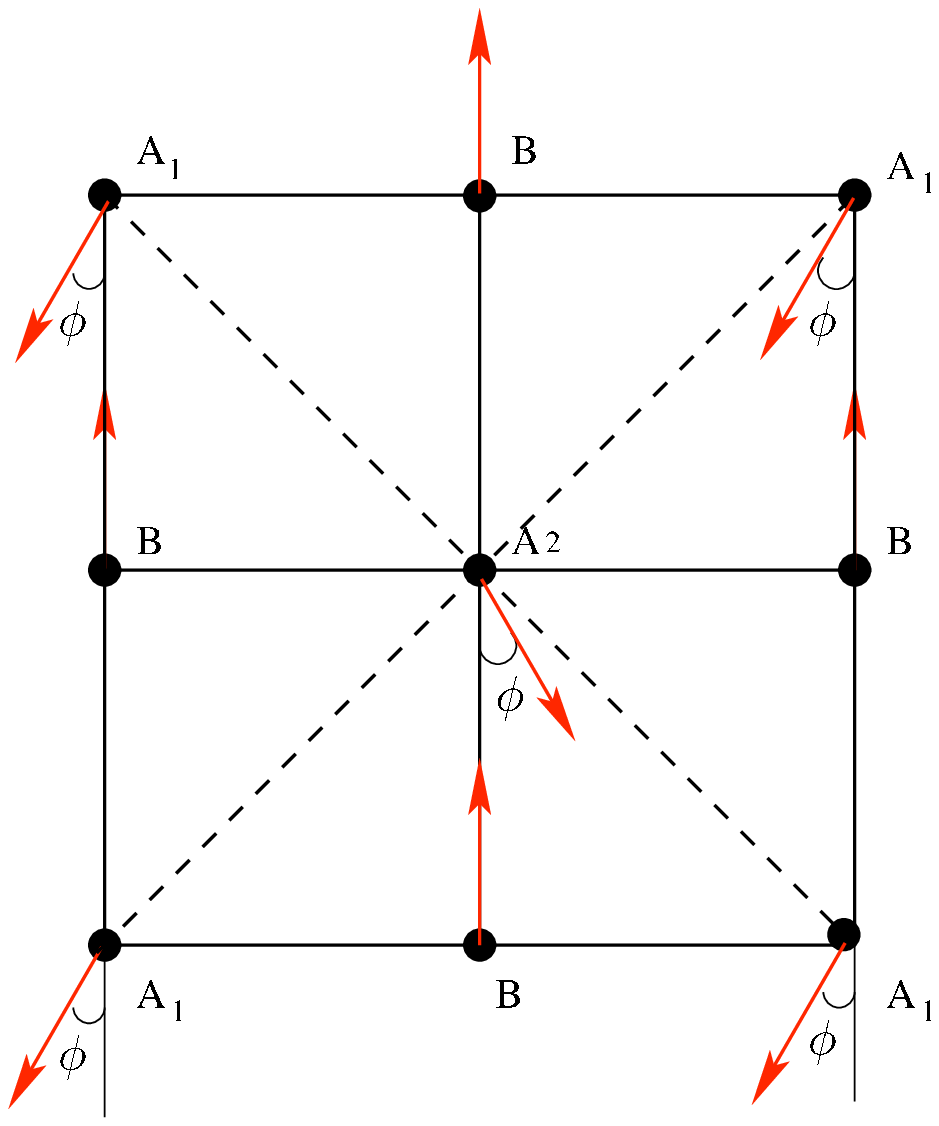}}}
   \subfloat[Semi-striped\label{SemiStripe_model}]{\scalebox{0.45}{\epsfig{file=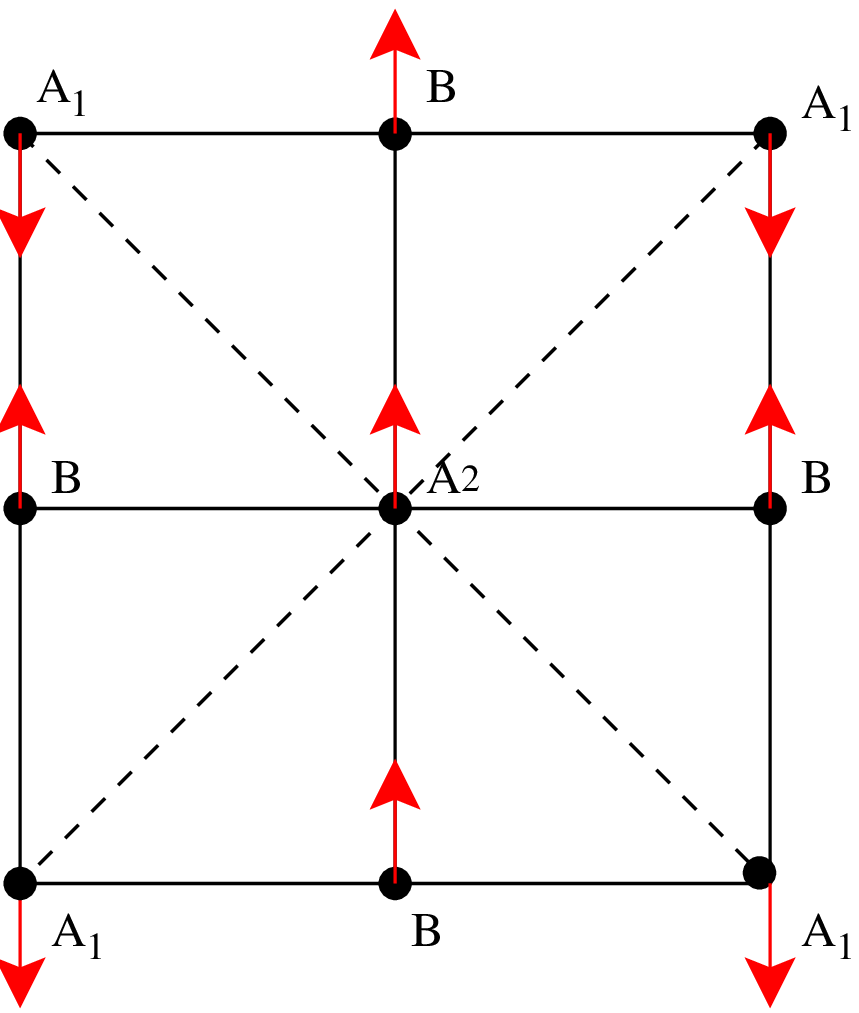}}}
}
\caption{(Color online) Union Jack model; --- $J_{1}$; - - -
  $J_{2}$. (a) Canted state; (b) Semi-striped state. The unit cell is a
  square of side length 2.}
\label{model}
\end{figure}

In this paper we now apply the CCM to the spin-half Union Jack model
that has been studied recently by other
means.~\cite{Co:2006_PRB,Co:2006_JPhys,Zh:2007} Its
Hamiltonian is written as
\begin{equation}
H = J_{1}\sum_{\langle i,j \rangle}{\bf s}_{i}\cdot{\bf s}_{j} + J_{2}\sum_{[i,k]}{\bf s}_{i}\cdot {\bf s}_{k}\,,  \label{H}
\end{equation}
where the operators ${\bf s}_{i} \equiv (s^{x}_{i}, s^{y}_{i},
s^{z}_{i})$ are the quantum spin operators on lattice site $i$ with ${\bf
  s}^{2}_{i} = s(s+1)$ and $s=1/2$. On the square lattice the sum over
$\langle i,j \rangle$ runs over all distinct NN bonds with strength
$J_{1}$, while the sum over [$i,k$] runs over only half of the
distinct NNN diagonal bonds having strength $J_{2}$ and with only one
diagonal bond on each square plaquette as arranged in the pattern
shown explicitly in Fig.\ \ref{model}.
The unit cell is thus the $2 \times 2$ square shown in Fig.\
\ref{canted_model}. (We note that, by contrast, the $J_{1}$--$J_{2}$
model discussed above, includes all of the diagonal NNN bonds on the
square lattice.) We consider here the case where both sorts of bonds
are antiferromagnetic, $J_{1} > 0$ and $J_{2} \equiv \kappa J_{1} >
0$, and are hence acting to compete against (or to frustrate) each
other. Henceforth we set $J_{1} \equiv 1$. We consider the model
equivalently defined by the Union Jack geometry in which there are two
sorts of sites, namely the A sites with 8 NN sites and the B sites
with 4 NN sites, as shown in Fig.\ \ref{canted_model}.

Considered classically rather than quantum-mechanically, (and thus
corresponsing to the quantum case in the limit where the spin quantum
number $s \rightarrow \infty$), the Union Jack model has only two gs
phases as the parameter $\kappa$ is varied over the range (0,
$\infty$). A simple variational analysis for the classical model
reveals that for $0 < \kappa$ $< 1/2$ the gs phase is
N\'{e}el-ordered, exactly as for the full $J_{1}$--$J_{2}$ model. Thus
the N\'{e}el ordering induced by the $J_{1}$-bonds acting alone is
preserved as the strength of the competing $J_{2}$-bonds is increased,
until the critical value $\kappa^{{\rm cl}}_{c}=0.5$ is reached. For
$\kappa > \kappa^{{\rm cl}}_{c}$ a new phase of lower energy emerges,
just as in the full $J_{1}$--$J_{2}$ model. However, whereas for the
full $J_{1}$--$J_{2}$ model that new phase is a classical striped
state in which alternate rows (or columns) of spins are arranged
antiparallel to one another, the new classical gs phase for the Union
Jack model is the canted ferrimagnetic state shown in Fig.\
\ref{canted_model} in which the spins on each of the alternating
A$_{1}$ and A$_{2}$ sites of the A-sublattice are canted respectively
at angles ($\pi \mp \phi$) with respect to those on the B-sublattice,
all of the latter of which point in the same direction. On the
A-sublattice each site A$_{1}$ has 4 NN sites A$_{2}$, and vice
versa. The angle between the NN spins on the A-sublattice is thus
2$\phi$.

The classical energy of the above canted state is thus 
\begin{equation}
E=Ns^{2}(\kappa\,{\rm cos}\,2\phi - 2\,{\rm cos}\,\phi)\,,
\end{equation}
where $J_{1} \equiv 1$ and $N \rightarrow \infty$ is the number of
sites. Clearly the energy is extremized when
\begin{equation}
{\rm sin}\,\phi(1 - 2 \kappa\, {\rm cos}\, \phi) = 0\,.
\end{equation}
When $\kappa < \kappa^{{\rm cl}}_{c} \equiv 0.5$, the lowest energy
corresponds to sin$\,\phi=0$ and hence to the N\'{e}el state. By
contrast, when $\kappa > \kappa^{{\rm cl}}_{c} \equiv 0.5$ the lowest
energy solution is the canted state with
\begin{equation}
\phi_{{\rm cl}} = {\rm cos}^{-1}\,\bigg(\frac{1}{2\kappa}\bigg) \,.  \label{pitch_angle}
\end{equation}
Thus the classical gs energy is given by
\begin{equation}
E^{{\rm cl}} = \left\{ 
\begin{array}{l l}
  Ns^{2}(\kappa -2)\,; & \quad \kappa < \kappa^{{\rm cl}}_{c} \equiv 0.5 \,.\\
  Ns^{2}\big(-\frac{1}{2\kappa} - \kappa\big)\,; & \quad \kappa > \kappa^{{\rm cl}}_{c} \equiv 0.5\,.\\ \end{array} \right. 
\end{equation}
The classical phase transition at $\kappa = \kappa^{{\rm cl}}_{c}
\equiv 0.5$ is of continuous (second-order) type with the gs energy
and its first derivative both continuous functions of $\kappa$,
although there are finite discontinuities in the second- and
higher-order derivatives at $\kappa = \kappa^{{\rm cl}}_{c}$.

In the classical canted phase the total magnetization per site is
$m^{{\rm cl}}=\frac{1}{2}s[1-(2\kappa)^{-1}]$, and the model thus exhibits
ferrimagnetism in this phase. Whereas ferrimagnetism more commonly
occurs when the individual ionic spins have different magnitudes on
different sublattices, it arises here in a case where when the spins
all have the same magnitude and all the interactions are
antiferromagnetic in nature, but the frustration between them acts to
produce an overall magnetization. The total magnetization $m$ vanishes
linearly as $\kappa \rightarrow \kappa^{{\rm cl}}_{c}$ from the canted
phase and then remains zero in the N\'{e}el phase for $\kappa <
\kappa$$^{{\rm cl}}_{c}$. The spontaneous breaking of the spin rotation
symmetry is also reflected by the vanishing of the energy gap on both
sides of the transition. Clearly on both sides of the transition the
translation symmetry of the lattice is also broken.

One of the aims of the present paper is to give a fully microscopic
analysis of the Union Jack model for the quantum case where the spins
all have spin quantum number $s=1/2$. We are interested to map out the
zero-temperature ($T=0$) phase diagram of the model, including the
positions and orders of any quantum phase transitions that emerge. In particular
we investigate the quantum analogs of the classical N\'{e}el and
canted phases and calculate the effect of quantum fluctuations on the
position and nature of the transition between them. We also aim to
investigate, for particular regions of the control parameter $\kappa$,
whether the quantum fluctuations may favor other phases, which have
no classical counterparts. One such possible candidate is discussed
below.

In the limit of $\kappa \rightarrow \infty$ the above classical limit
corresponds to a canting angle $\phi \rightarrow \frac{1}{2} \pi$,
such that the spins on the A-sublattice become N\'{e}el-ordered, as is
expected. The spins on the antiferromagnetically-ordered A-sublattice
are orientated at 90$^{\circ}$ to those on the
ferromagnetically-ordered B-sublattice in this limit. In reality, of
course, there is complete degeneracy at the classical level in this
limit between all states for which the relative ordering directions
for spins on the A- and B- sublattices are arbitrary. Clearly the
exact spin-1/2 limit should also comprise decoupled antiferromagnetic and
ferromagnetic sublattices. However, one might now expect that this
degeneracy in the relative spin orientations between the two
sublattices is lifted by quantum fluctuations by the well-known
phenomenon of {\it order by disorder}.\cite{Vi:1977} Just such a phase
is known to exist in the full spin-1/2 $J_{1}$--$J_{2}$ model for
values of $J_{2}/J_{1} \gtrsim 0.6$, where it is the so-called
collinear striped phase in which, on the square lattice, spins along
(say) the rows in Fig.\ \ref{model} order ferromagnetically while
spins along the columns and diagonals order antiferromagnetically. We
have also shown how such a striped state is stabilized by quantum
fluctuations for values of $J_{2}'/J_{1} \gtrsim 1.8$ for the spin-1/2
$J_{1}$--$J_{2}'$ model defined on an anistropic 2D
lattice,\cite{square_triangle} as discussed in Sec.\ \ref{Introd} above.

The existence of the striped state as a stable phase for large values
of the frustration parameter for both the spin-1/2 $J_{1}$--$J_{2}$
and $J_{1}$--$J_{2}'$ models above is a reflection of the well-known
fact that quantum fluctuations favor collinear ordering. In both
cases the order-by-disorder mechanism favors the collinear state from
the otherwise infinitely degenerate set of available states at the
classical level. For the present Union Jack model the corresponding
collinear state that might perhaps be favored by the order by
disorder mechanism is the so-called semi-striped state shown in Fig.\
\ref{SemiStripe_model} where the A-sublattice is now N\'{e}el-ordered
in the same direction as the B-sublattice is ferromagnetically
ordered. Alternate rows (or columns) are thus ferromagnetically and
antiferromagnetically ordered in the same direction. We investigate
the possibility below that such a semi-stripe-ordered phase may be
stabilized by quantum fluctuations at larger values of $\kappa$.

\section{The coupled cluster method}
The CCM (see, e.g., Refs.~[\onlinecite{Bi:1991,Bi:1998,Fa:2004}] and
references cited therein) that we employ here is one of the most
powerful and most versatile modern techniques available to us in
quantum many-body theory. It has been applied very successfully to
various quantum magnets (see
Refs.~[\onlinecite{j1j2_square_ccm1,Bi:2008_PRB,Bi:2008_JPCM,j1j2_square_ccm4,j1j2_square_ccm5,Fa:2001,Kr:2000,Schm:2006,Fa:2004,Ze:1998,shastry2,shastry3,square_triangle}]
and references cited therein). The method is particularly appropriate
for studying frustrated systems, for which some of the main
alternative methods either cannot be applied or are sometimes only of
limited usefulness, as explained below. For example, QMC techniques
are particularly plagued by the sign problem for such systems, and the
ED method is restricted in practice by available computational power,
particularly for $s>1/2$, to such small lattices that it is often
insensitive to the details of any subtle phase order present.

The method of applying the CCM to quantum magnets has been described
in detail elsewhere (see, e.g.,
Refs.~[\onlinecite{Fa:2001,Kr:2000,Schm:2006,Bi:1991,Bi:1998,Fa:2004}]
and references cited therein). It relies on building multispin
correlations on top of a chosen gs model state $|\Phi\rangle$ in a
systematic hierarchy of LSUB$n$ approximations (described below) for
the correlation operators $S$ and $\tilde{S}$ that parametrize the
exact gs ket and bra wave functions of the system respectively as
$|\Psi \rangle=e^{S}|\Phi\rangle$ and $\langle \tilde{\Psi}|=\langle
\Phi|\tilde{S}e^{-S}$. In the present case we use three different
choices for the model state $|\Phi\rangle$, namely either of the
classical N\'{e}el and canted states, as well as the semi-striped
state. Note that for the canted phase we perform calculations for
arbitrary canting angle $\phi$ [as shown in Fig.\ \ref{canted_model}],
and then minimize the corresponding LSUB$n$ approximation for the
energy with respect to $\phi$, $E_{{\rm LSUB}n}(\phi) \rightarrow {\rm
  min} \Leftrightarrow \phi = \phi_{{\rm LSUB}n}$. Generally (for $n >
2$) the minimization must be carried out computationally in an
iterative procedure, and for the highest values of $n$ that we use
here the use of supercomputing resources was essential. Results for
the canting angle $\phi_{{\rm LSUB}n}$ will be given later. We choose
local spin coordinates on each site in each case so that all spins in
$|\Phi\rangle$, whatever the choice, point in the negative
$z$-direction (i.e., downwards) by definition in these local
coordinates.

Then, in the LSUB$n$ approximation all possible multi-spin-flip
correlations over different locales on the lattice defined by $n$ or
fewer contiguous lattice sites are retained. The numbers of such
distinct (i.e., under the symmetries of the lattice and the model
state) fundamental configurations of the current model in various
LSUB$n$ approximations are shown in Table~\ref{table_FundConfig}.
\begin{table}[t]
  \caption{Number of fundamental LSUB$n$ configurations ($N_{f}$) for the semi-striped and 
    canted states of the spin-$1/2$ Union Jack model, based on the Union Jack geometry defined in the text.}
\label{table_FundConfig}
\begin{tabular}{cccc} \hline\hline
\multirow{2}{*}{Method} &  \multicolumn{3}{c}{$N_{f}$} \\ \cline{2-4}
& semi-striped & & canted \\ \hline
LSUB$2$ & 3    & & 5      \\ 
LSUB$3$ & 5    & & 42     \\ 
LSUB$4$ & 41   & & 199    \\ 
LSUB$5$ & 194  & & 1259   \\ 
LSUB$6$ & 1159 & & 8047   \\ 
LSUB$7$ & 6862 & & 56442  \\ \hline\hline
\end{tabular} 
\end{table}
We note that the distinct configurations given in Table
\ref{table_FundConfig} are defined with respect to the Union Jack
geometry described in Sec.\ \ref{model_section}, in which the
B-sublattice sites of Fig.\ \ref{canted_model} are defined to have 4
NN sites and the A-sublattice sites are defined to have the 8 NN sites
joined to them either by $J_{1}$- or $J_{2}$-bonds. If we chose instead
to work in the square-lattice geometry every site would have 4 NN
sites. The coupled sets of equations for these corresponding numbers
of coefficients in the operators $S$ and $\tilde{S}$ are derived using
computer algebra~\cite{ccm} and then solved~\cite{ccm} using parallel
computing. We note that such CCM calculations using up to about
$10^{5}$ fundamental configurations or so have been previously carried
out many times using the CCCM code\cite{ccm} and heavy
parallelization. A significant extra computational burden arises here
for the canted state due to the need to optimize the quantum canting
angle $\phi$ at each LSUB$n$ level of approximation as described
above. Furthermore, for many model states the quantum number
$s^{z}_{T} \equiv \sum^{N}_{i=1} s^{z}_{i}$ in the original global
spin-coordinate frame, may be used to restrict the numbers of
fundamental multi-spin-flip configurations to those clusters that
preserve $s^{z}_{T}$ as a good quantum number. This is true for the
N\'{e}el state where $s^{z}_{T} = 0$ and for the semi-striped state
for which $s^{z}_{T} = N/4$, where $N$ is the number of lattice
sites. However, for the canted model state that symmetry is absent,
which largely explains the significantly greater number of fundamental
configurations shown in Table \ref{table_FundConfig} for the canted
state at a given LSUB$n$ order. Hence, the maximum LSUB$n$ level that
we can reach here for the canted state, even with massive
parallelization and the use of supercomputing resources, is
LSUB$7$. For example, to obtain a single data point for a given value
of $\kappa$ (i.e., for a given value of $J_{2}$, with $J_{1}=1$) for
the canted phase at the LSUB$7$ level typically required about 0.3
hours computing time using 600 processors simultaneously. However, for
values of $\kappa$ near to termination points at which CCM solutions
using that model state disappear (as described more fully below), the
computing time typically increased significantly.

At each level of approximation we may then calculate a corresponding
estimate of the gs expectation value of any physical observable such
as the energy $E$ and the magnetic order parameter, $M \equiv
-\frac{1}{N}\sum^{N}_{i=1}\langle \tilde{\Psi}|s^{z}_{i}|\Psi
\rangle$, defined in the local, rotated spin axes, and which thus
represents the average on-site magnetization. Note that $M$
is just the usual sublattice (or staggered) magnetization per site for
the case of the N\'{e}el state as the CCM model state, for example.

It is important to note that we never need to perform any finite-size
scaling, since all CCM approximations are automatically performed from
the outset in the infinite-lattice limit, $N \rightarrow \infty$,
where $N$ is the number of lattice sites. However, we do need as a
last step to extrapolate to the $n \rightarrow \infty$ limit in the
LSUB$n$ truncation index $n$. We use here the
well-tested~\cite{Fa:2001,Kr:2000} empirical scaling laws
\begin{equation}
E/N=a_{0}+a_{1}n^{-2}+a_{2}n^{-4}\,,  \label{Extrapo_E}
\end{equation} 
\begin{equation}
M=b_{0}+b_{1}n^{-1}+b_{2}n^{-2}\,. \label{Extrapo_M}
\end{equation} 

\section{RESULTS}

We report here on CCM calculations for the present spin-1/2 Union Jack
model Hamiltonian of Eq.\ (\ref{H}) for given parameters ($J_{1}=1$,
$J_{2}$), based respectively on the N\'{e}el, canted and semi-striped
states as CCM model states. Our computational power is such that we
can perform LSUB$n$ calculations for each model state with $n \leq
7$. We note that, as has been well documented in the
past,~\cite{Fa:2008} the LSUB$n$ data for both the gs energy per spin
$E/N$ and the average on-site magnetization $M$ converge differently
for the even-$n$ and the odd-$n$ sequences, similar to what is
frequently observed in perturbation theory.~\cite{Mo:1953} Since, as a
general rule, it is desirable to have at least ($n+1$) data points to
fit to any fitting formula that contains $n$ unknown parameters, we
prefer to have at least 4 results for different values of the LSUB$n$
truncation index $n$ to fit to Eqs.\ (\ref{Extrapo_E}) and
(\ref{Extrapo_M}). However, for all of our extrapolated results below
we perform separate extrapolations using the even and odd LSUB$n$
sequences with $n=\{2,4,6\}$ and $n=\{3,5,7\}$.

\subsection{N\'{e}el state versus the canted state}
\begin{figure*}[t]
\mbox{
  \subfloat[\label{EvsAngle_general}]{\scalebox{0.3}{\epsfig{file=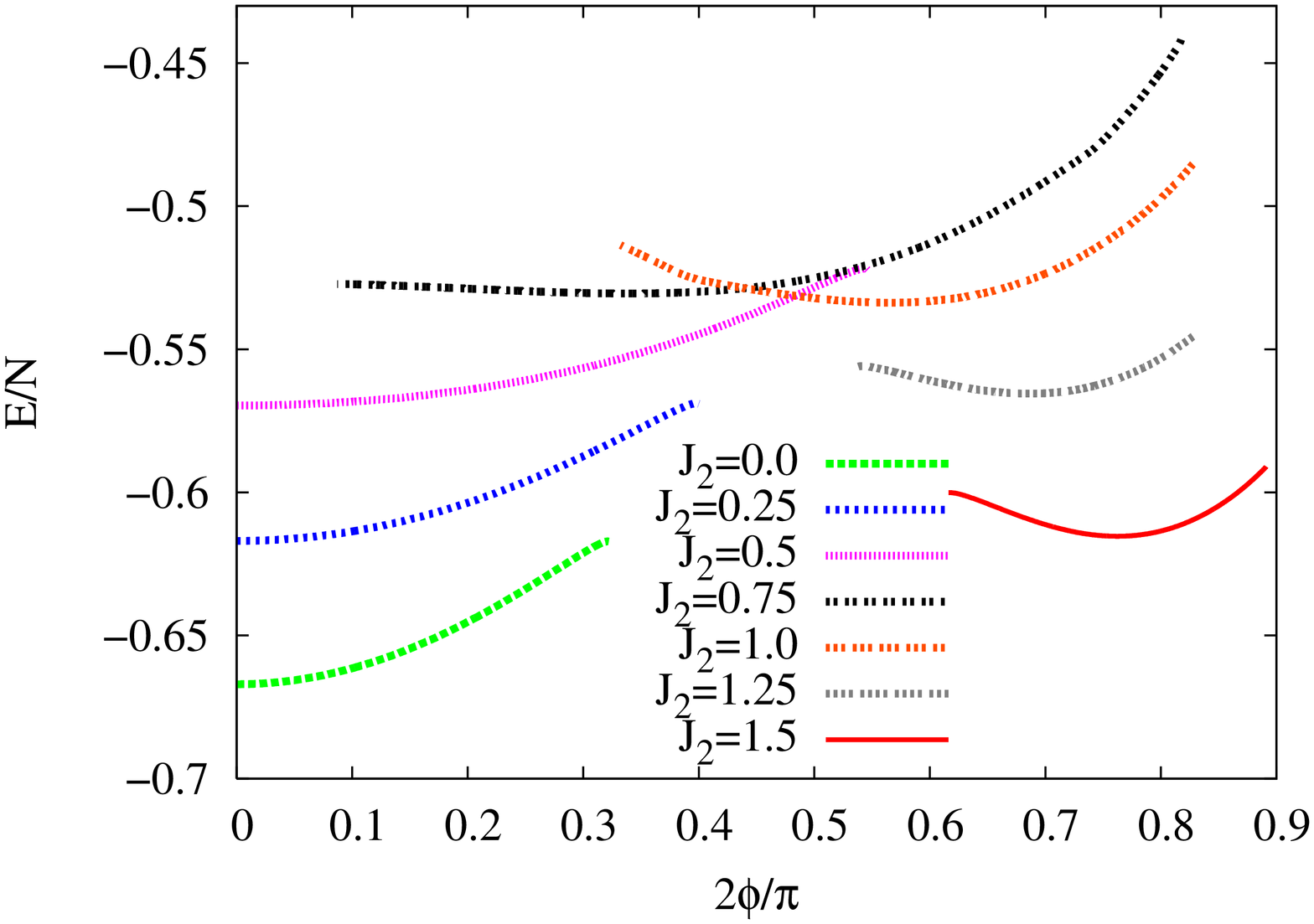,angle=270}}}
}
\mbox{
  \subfloat[\label{EvsAngle_detailed}]{\scalebox{0.3}{\epsfig{file=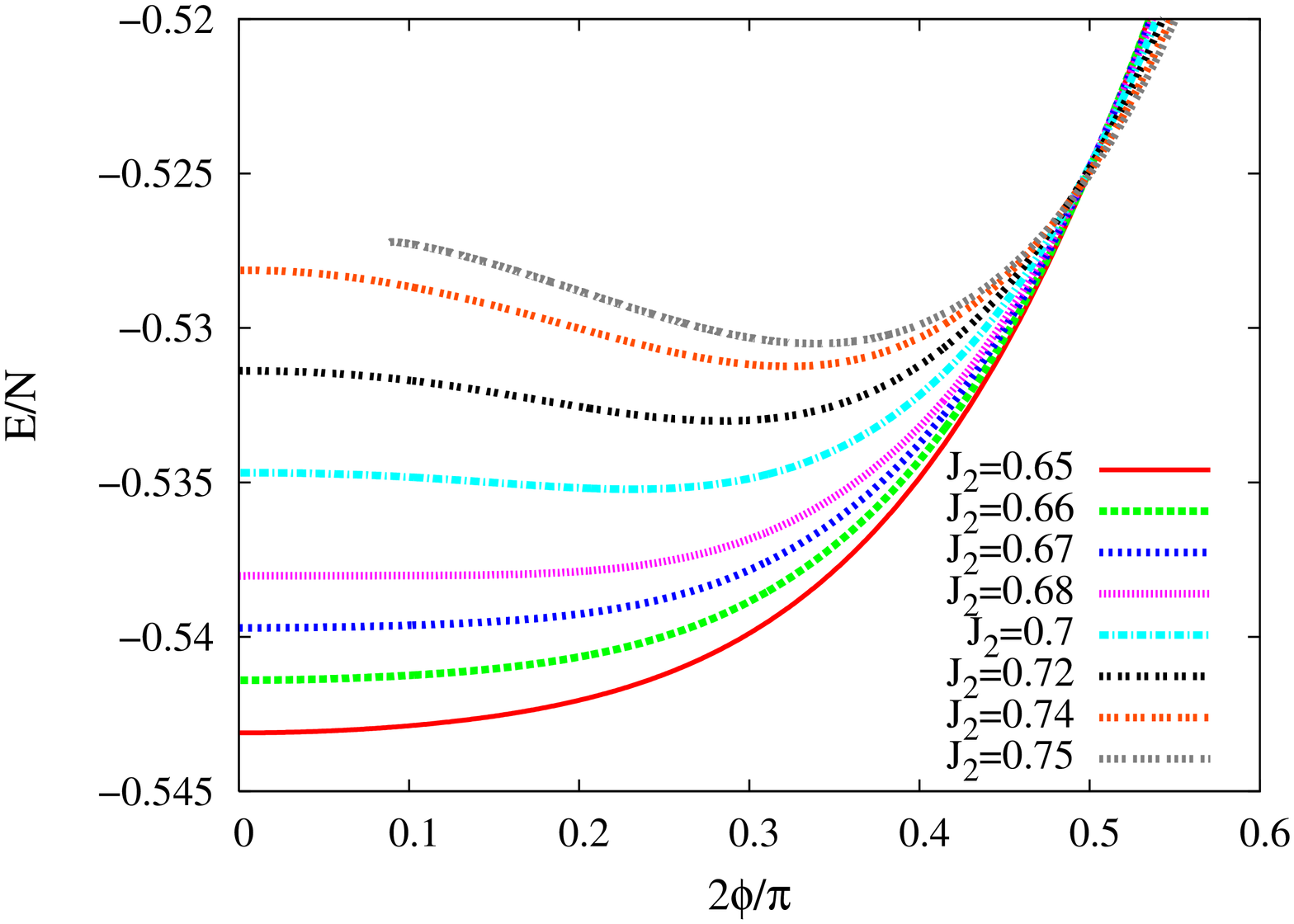,angle=270}}}
}
\caption{(Color online) Ground-state energy per spin of the spin-1/2
  Union Jack Hamiltonian of Eq.\ (\ref{H}) with $J_{1} \equiv 1$,
  using the LSUB6 approximation of the CCM with the canted model
  state, versus the canting angle $\phi$, for some illustrative values
  of $J_{2}$ in the range $0 \leq J_{2} \leq 1.5$ for Fig.\
  \ref{EvsAngle_general} and $0.65 \leq J_{2} \leq 0.75$ for Fig.\
  \ref{EvsAngle_detailed}. For $J_{2} \lesssim 0.68$ in this
  approximation the minimum is at $\phi=0$ (N\'{e}el order), whereas
  for $J_{2} \gtrsim 0.68$ the minimum occurs at $\phi=\phi_{{\rm
      LSUB}6} \neq 0$, indicating a phase transition at $J_{2} \approx
  0.68$ in this LSUB6 approximation.}
\label{EvsAngle}
\end{figure*}

\begin{figure*}[t]
\mbox{
  \subfloat[$n=\{2,4,6\}$]{\scalebox{0.3}{\epsfig{file=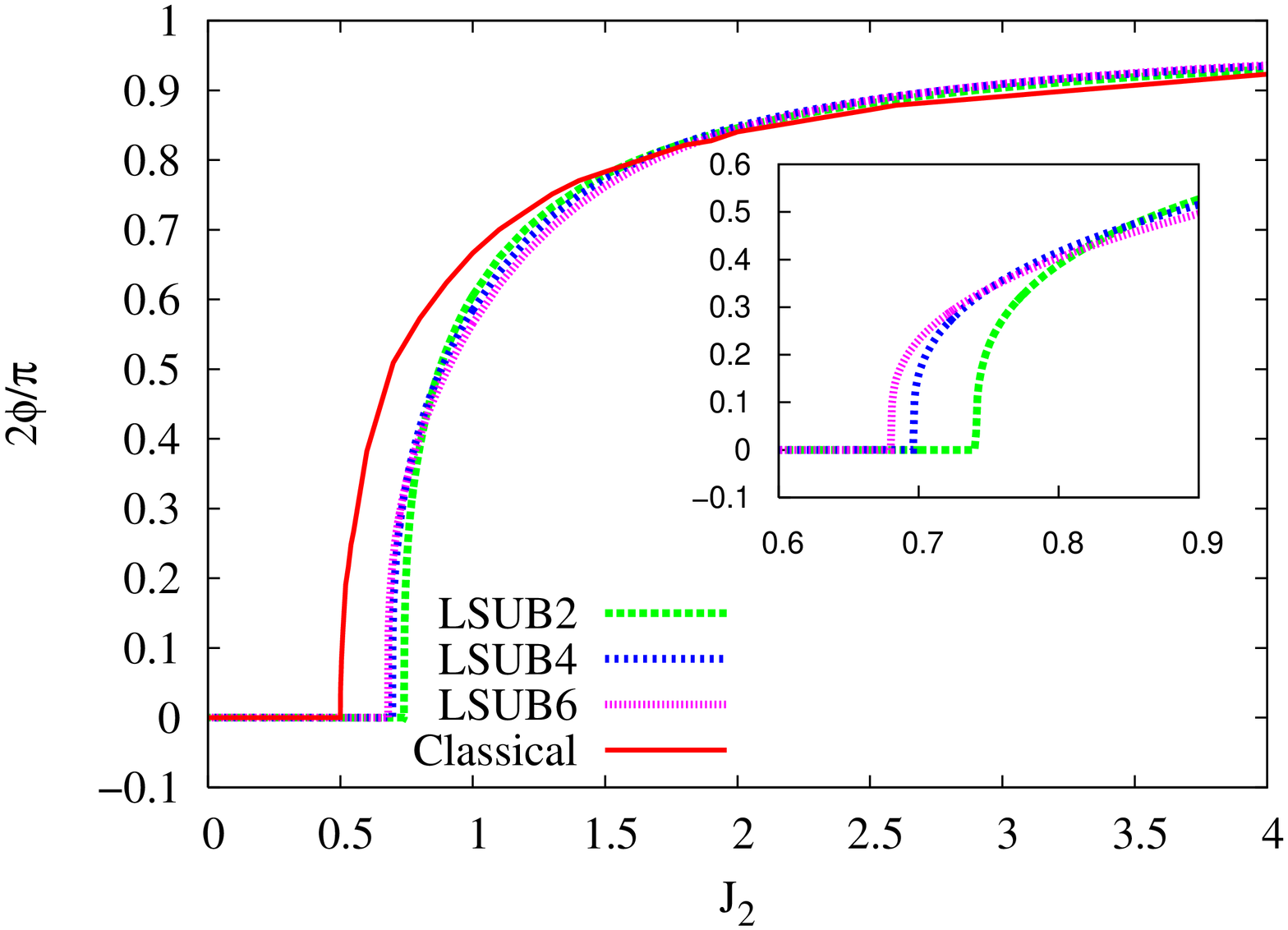,angle=270}}}
}
\mbox{
  \subfloat[$n=\{3,5,7\}$]{\scalebox{0.3}{\epsfig{file=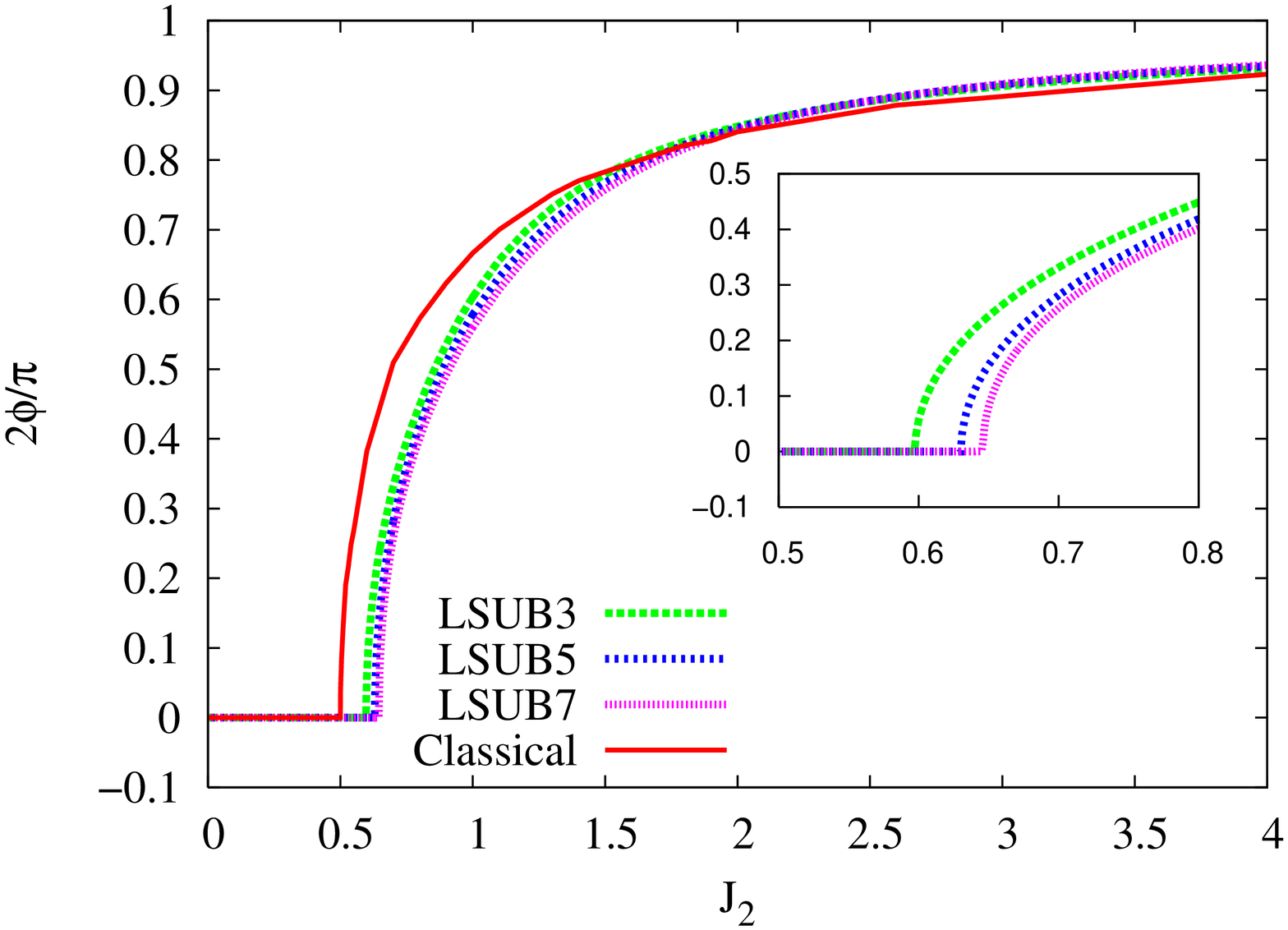,angle=270}}}
}
\caption{(Color online) The angle $\phi_{{\rm LSUB}n}$ that minimizes
  the energy $E_{{\rm LSUB}n}(\phi)$ of the spin-1/2 Union Jack
  Hamiltonian of Eq.\ (\ref{H}) with $J_{1} \equiv 1$, in the LSUB$n$
  approximations with (a) $n=\{2,4,6\}$ and (b) $n=\{3,5,7\}$, using the
  canted model state, versus $J_{2}$. The corresponding classical
  result $\phi_{{\rm cl}}$ from Eq.~(\ref{pitch_angle}) is shown for
  comparison. We find in the LSUB$n$ quantum case with $n > 2$ a
  weakly first-order phase transition or second-order phase transition
  (e.g., for LSUB6 at $J_{2} \approx 0.680$ and LSUB7 at $J_{2}
  \approx 0.646$). By contrast, in the classical case there is a
  second-order phase transition at $J_{2}=0.5$.}
\label{angleVSj2}
\end{figure*}

We report first on results obtained using the N\'{e}el and canted
model states.  While classically we have a second-order phase
transition from N\'{e}el order (for $\kappa < \kappa$$^{{\rm cl}}_{c}$) to
canted order (for $\kappa > \kappa^{{\rm cl}}_{c}$), where $\kappa
\equiv J_{2}/J_{1}$, at a value $\kappa^{{\rm cl}}_{c} = 0.5$, using
the CCM we find strong indications of a shift of this critical point
to a higher value $\kappa_{c_{1}} \approx 0.66$ in the spin-1/2
quantum case as we explain in detail below. Thus, for example, curves
such as those shown in Fig.\ \ref{EvsAngle}
show that the N\'{e}el model state ($\phi=0$) gives the minimum gs
energy for all values of $\kappa < \kappa_{c_{1}}$ where
$\kappa_{c_{1}} = \kappa^{{\rm LSUB}n}_{c_{1}}$ is also dependent on
the level of LSUB$n$ approximation, as we see clearly in Fig.\
\ref{angleVSj2}.

By contrast, for $\kappa > \kappa_{c_{1}}$ the minimum in the energy
is found to occur at a value $\phi \neq 0$. If we consider the canting
angle $\phi$ itself as an order parameter (i.e., $\phi=0$ for N\'{e}el
order and $\phi \neq 0$ for canted order) a typical scenario for a
first-order phase transition would be the appearance of a two-minimum
structure for the gs energy as a function of $\phi$. If we therefore
admit such a scenario, in the typical case one would expect various
special points in the transition region, namely the phase transition
point $\kappa_{c_{1}}$ itself where the two minima have equal depth,
plus one or two instability points $\kappa_{i_{1}}$ and
$\kappa_{i_{2}}$ where one or other of the minima (at $\phi = 0$ and
$\phi \neq 0$ respectively) disappears. By contrast, a second-order
phase transition might manifest itself via a one-minimum structure for
the gs energy as a function of $\phi$, in which the single minimum
moves smoothly and continuously from the value $\phi = 0$ for all
values of $\kappa < \kappa_{c_{1}}$ to nonzero value $\phi \neq 0$ for
$\kappa > \kappa_{c_{1}}$. 

We show in Fig.\ \ref{EvsAngle} our results
for the LSUB6 approximation based on the canted (or N\'{e}el) state
as the CCM model state. Very similar curves occur for other LSUB$n$
approximations. A close inspection of curves such as those shown in
Fig.\ \ref{EvsAngle} for the LSUB6 case shows that what happens for
this model at this level of approximation is that for $\kappa \lesssim
0.68$ the only minimum in the gs energy is at $\phi = 0$ (N\'{e}el
order). As this value is approached from below the LSUB6 energy curves
become extremely flat near $\phi = 0$, indicating the disappearance at
$\phi = 0$ of the second derivative $d^{2}E/d\phi^{2}$ (and possibly
also of one or more of the higher derivatives $d^{n}E/d\phi^{n}$ with
$n \geq$ 3), as well as of the first derivative $dE/d\phi$. Then,
for all values $\kappa \gtrsim 0.68$ the LSUB6 curves develop a
minimum at a value $\phi \neq 0$ which is also the global minimum. The
state for $\phi \neq 0$ is thus the quantum analog of the classical
canted phase. The fact that the antiferromagnetic N\'{e}el order
survives into the classically unstable regime is another example of
the well-known phenomenon that quantum fluctuations tend to promote
collinear order in magnetic spin-lattice systems, as has been observed
in many other such cases (see e.g.,
Ref.~[\onlinecite{Kr:2000,Ko:1996}]). Thus, this collinear
N\'{e}el-ordered state survives into a region where classically it
becomes unstable with respect to the non-collinear canted state.

A close inspection of the curves shown in Fig.~\ref{angleVSj2} for
various LSUB$n$ approximation shows that the crossover from one
minimum ($\phi = 0$, N\'{e}el) solution to the other ($\phi \neq 0$,
canted) appears to be continuous for the odd-$n$ sequence, thus
indicating a second-order transition according to the above
scenario. By contrast, for the even-$n$ sequence with $n > 2$ the
curves in Fig.~\ref{angleVSj2} become very steep in the crossover
region just above $\kappa^{{\rm LSUB}n}_{c_{1}}$ and due to the
extremely flat nature of the gs energy curves as a function of $\phi$
in this region, as shown in Fig.~\ref{EvsAngle}, it is impossible to
rule out a small but finite discontinuity in the curves of
Fig.~\ref{EvsAngle_general} for the even-$n$ LSUB$n$ sequence at
$\kappa = \kappa^{{\rm LSUB}n}_{c_{1}}$. However, if the phase
transition is, in fact first-order, it is certainly only very weakly
so according to this criterion.

Thus, based on the evidence presented so far of the gs energies of the
N\'{e}el and canted phases, it would appear that the transiton at
$\kappa = \kappa_{c_{1}}$ between these two phases is either
second-order, as in the classical phase, or weakly first-order.  Such
a situation where the quantum fluctuations change the nature of a
phase transition qualitatively from a classical second-order type to a
quantum first-order type has also been seen previously in the
comparable spin-1/2 HAF models that interpolate continuously between
the square and triangular lattices,~\cite{square_triangle} and between
the square and honeycomb lattices,~\cite{Kr:2000} respectively. In the
present spin-1/2 Union Jack model, however, the CCM gs energy results
appear to favor a second-order transition, although the extreme
insensitivity of the gs energy to the canting angle $\phi$ near the
crossover region, especially for the even-$n$ LSUB$n$ sequence with $n
> 2$, means that we cannot rule out a weakly first-order
transition. The evidence to date indicates, however, that the quantum
phase transition at $\kappa_{c_{1}}$ is a subtle one. Furthermore, the
present spin-1/2 Union Jack model appears, on the evidence to date, to
behave somewhat differently (viz., in some senses ``more
classically'') than its corresponding spin-1/2 interpolating
square-triangle Heisenberg antiferromagnet
counterpart.\cite{square_triangle} Further evidence from
Fig.~\ref{angleVSj2} appears to back up this observation. Thus, we see
from Fig.~\ref{angleVSj2} that the quantum canting angle $\phi$
approaches its asymptotic value $\pi/2$ as $\kappa \rightarrow \infty$
slightly faster than does the corresponding classical value. By
contrast, in the case of the spin-1/2 interpolating square-triangle
Heisenberg antiferromagnet,\cite{square_triangle} the corresponding
pitch angle $\phi$ of the spiral phase (that is the analog of the
canted phase for the present model) approaches its similar asymptotic
value $\pi/2$ as $\kappa \rightarrow \infty$ {\it very much} faster
than does the classical value. We also discuss this difference more
fully below, where we find further evidence that quantum fluctuations
modify the classical behavior of the Union Jack model rather less than
they do for its corresponding spin-1/2 interpolating square-triangle
Heisenberg antiferromagnet counterpart.

We show in Table~\ref{table_CritPt}
\begin{table}[t]
  \caption{The critical value $\kappa^{{\rm LSUB}n}_{c_{1}}$ at which the transition between the N\'{e}el phase ($\phi=0$) and the canted phase ($\phi \neq 0$) occurs in the LSUB$n$ approximation using the CCM with (N\'{e}el or) canted state as model state.}
\label{table_CritPt}
\begin{tabular}{cc} \hline\hline
Method & $\kappa^{{\rm LSUB}n}_{c_{1}}$ \\ \hline
LSUB$2$ & 0.740 \\ 
LSUB$4$ & 0.696 \\ 
LSUB$6$ & 0.680 \\ 
LSUB$\infty$ $^{a}$ & 0.651 $\pm 0.001$\,. \\ 
LSUB$\infty$ $^{b}$ & 0.676 $\pm 0.004$\,. \\ \hline
LSUB$3$ & 0.597  \\ 
LSUB$5$ & 0.630   \\ 
LSUB$7$ & 0.645  \\ 
LSUB$\infty$ $^{a}$ & 0.681 $\pm 0.001$\,. \\ 
LSUB$\infty$ $^{b}$ & 0.653 $\pm 0.004$\,. \\ \hline\hline
\end{tabular}
\vspace{0.3cm}
\\ \protect $^{a}$ Based on $\kappa^{{\rm LSUB}n}_{c_{1}}=a_{0}+a_{1}n^{-1}$, with $n=\{2,4,6\}$ or $n=\{3,5,7\}$\,. \\
\protect $^{b}$ Based on $\kappa^{{\rm LSUB}n}_{c_{1}}=b_{0}+b_{1}n^{-2}$, with $n=\{2,4,6\}$ or $n=\{3,5,7\}$\,.
\end{table}
the critical values $\kappa^{{\rm LSUB}n}_{c_{1}}$ at which the
transition between the N\'{e}el and canted phases occurs in the
various LSUB$n$ approximations shown in Fig.~\ref{angleVSj2}. In the
past we have found that a simple linear extrapolation, $\kappa^{{\rm
    LSUB}n}_{c_{1}} = a_{0}+a_{1}n^{-1}$, yields a good fit to such
critical points, as seems to be the case here too. The corresponding
``LSUB$\infty$'' estimates from the LSUB$n$ data in
Table~\ref{table_CritPt} are $\kappa_{c_{1}}=0.651 \pm 0.001$ based on
$n=\{2,4,6\}$ and $\kappa_{c_{1}}=0.681 \pm 0.001$ based on
$n=\{3,5,7\}$ where the quoted errors are simply the standard
deviations from the two fits. Similar estimates based on an
extrapolation $\kappa^{{\rm LSUB}n}_{c_{1}}=b_{0}+b_{1}n^{-2}$ are
also shown in Table~\ref{table_CritPt}, for which the standard
deviations are clearly greater. The fact that the two estimates based
on the even-$n$ and odd-$n$ LSUB$n$ sequences differ slightly from one
another is a reflection of the extreme insensitivity of the gs energy
to the canting angle $\phi$ near $\kappa^{{\rm LSUB}n}_{c_{1}}$, and
the difference between the two estimates is a rough indication of our
real error bars on $\kappa_{c_{1}}$. We also present other independent
estimates of $\kappa_{c_{1}}$ below.

We note from Fig.\ \ref{EvsAngle} that for certain values of $J_{2}$
with $J_{1} \equiv 1$ (or, equivalently, $\kappa$) CCM solutions at a
given LSUB$n$ level of approximation (viz., LSUB$6$ in Fig.\
\ref{EvsAngle}) exist only for certain ranges of the canting angle
$\phi$. For example, for the pure square-lattice HAF ($\kappa=0$) the
CCM LSUB$6$ solution based on a canted model state only exists for $0
\leq \phi \lesssim 0.161 \pi$.  In this case, where the N\'{e}el
solution is the stable ground state, if we attempt to move too far
away from N\'{e}el collinearity the CCM equations themselves become
``unstable'' and simply do not have a real solution. Similarly, we see
from Fig.\ \ref{EvsAngle} that for $\kappa=1.5$ the CCM LSUB$6$
solution exists only for $0.308 \pi \lesssim \phi \leq 0.445 \pi$. In
this case the stable ground state is a canted phase, and now if we
attempt either to move too close to N\'{e}el collinearity or to
increase the canting angle too close to its asymptotic value of
$\pi/2$, the real solution terminates.

Such terminations of CCM solutions are very common and are very well
documented.\cite{Fa:2004} In all such cases a termination point always
arises due to the solution of the CCM equations becoming complex at
this point, beyond which there exist two branches of entirely
unphysical complex conjugate solutions.\cite{Fa:2004} In the region
where the solution reflecting the true physical solution is real there
actually also exists another (unstable) real solution. However, only
the shown branch of these two solutions reflects the true
(stable) physical ground state, whereas the other branch does not.
The physical branch is usually easily identified in practice as the
one which becomes exact in some known (e.g., perturbative) limit.
This physical branch then meets the corresponding unphysical branch at
some termination point (with infinite slope on Fig.\ \ref{EvsAngle})
beyond which no real solutions exist.  The LSUB$n$ termination points
are themselves also reflections of the quantum phase transitions in
the real system, and may be used to estimate the position of the phase
boundary,\cite{Fa:2004} although we do not do so for this first
critical point since we have more accurate criteria discussed above as
well as below.

Before doing so, however, we wish to give some further indication of
the accuracy of our results. Thus in Table~\ref{table_EandM_results}
\begin{table*}[t]
  \caption{Ground-state energy per spin $E/N$ and magnetic order parameter $M$ (i.e., the average on-site magnetization) for the spin-1/2 square-lattice HAF. We show CCM results obtained for the Union Jack model with $J_{1}=1$ and $J_{2}=0$ using the N\'{e}el model state in various CCM LSUB$n$ approximations defined on the Union Jack geometry described in Sec.~\ref{model_section}. We compare our extrapolated ($n \rightarrow \infty$) results using Eqs.~(\ref{Extrapo_E}) and (\ref{Extrapo_M}) with the odd-$n$ and even-$n$ LSUB$n$ data sets with other calculations.}
\begin{tabular}{ccc} \hline\hline
Method  &  $E/N$ & $M$   \\  \hline
LSUB$2$ & -0.64833 &  0.4207 \\ 
LSUB$3$ & -0.65044 &  0.4151  \\      
LSUB$4$ & -0.66366 &  0.3821  \\    
LSUB$5$ & -0.66398 &  0.3795   \\      
LSUB$6$ & -0.66703 &  0.3630  \\             
LSUB$7$ & -0.66724 &  0.3606  \\  \hline\hline           
\multicolumn{3}{c}{Extrapolations} \\ \hline
LSUB$\infty$ $^{a}$  & -0.6698 & 0.316  \\ 
LSUB$\infty$ $^{b}$  & -0.6704 & 0.304  \\    
QMC $^{c}$   & -0.669437(5) &  0.3070(3)  \\ 
SE $^{d}$ &  -0.6693(1) & 0.307(1)   \\ \hline\hline
\end{tabular} 
\vspace{0.3cm}
\\ 
\protect $^{a}$ Based on $n=\{2,4,6\}$  \\ 
\protect $^{b}$ Based on $n=\{3,5,7\}$ \\ 
\protect $^{c}$ QMC (Quantum Monte Carlo) for square lattice\cite{Sa:1997}  \\ 
\protect $^{d}$ SE (Series Expansion) for square lattice\cite{square_series} 
\label{table_EandM_results}
\end{table*}
we show data for the case of the spin-1/2 HAF on the square lattice
(corresponding to the case $\kappa=0$ of the present Union Jack
model). We present our CCM results in various LSUB$n$ approximations
(with $2 \leq n \leq 7$) based on the Union Jack geometry using the
N\'{e}el model state. Results are given for the gs energy per spin
$E/N$, and the magnetic order parameter $M$. We also display our
extrapolated ($n \rightarrow \infty$) results using the schemes of
Eqs.~(\ref{Extrapo_E}) and (\ref{Extrapo_M}) with the data sets
$n=\{2,4,6\}$ and $n=\{3,5,7\}$. The results are clearly seen to be
robust and consistent, and for comparison purposes we also show the
corresponding results using a QMC technique\cite{Sa:1997} and from a
linked-cluster series expansion (SE) method.\cite{square_series} We note
that for the square-lattice HAF no dynamic (or geometric) frustration
exists and the Marshall-Peierls sign rule\cite{Ma:1955} applies and
may be used to circumvent the QMC ``minus-sign problem''. The QMC
results\cite{Sa:1997} are thus extremely accurate for this limiting
($\kappa=0$) case only, and represent the best available results in
this case. Our own extrapolated results are in good agreement with
these QMC benchmark results, as found previously (see, e.g.,
Ref.~[\onlinecite{Fa:2008}] and references cited therein) for CCM
calculations performed specifically using the square lattice geometry,
as well as for other CCM calculations for which the square-lattice HAF
is a limit, such as for the spin-1/2 interpolating
square-triangle $J_{1}$--$J_{2}'$ model,\cite{square_triangle} for which the triangular lattice
geometry was employed. It is gratifying to note in particular that
although the individual LSUB$n$ results for the spin-1/2
square-lattice HAF depend upon which geometry is used to define the
configurations, the corresponding LSUB$\infty$ extrapolations are in
excellent agreement with one another.

\begin{figure*}[t]
\mbox{
  \subfloat[$n=\{2,4,6\}$]{\scalebox{0.3}{\epsfig{file=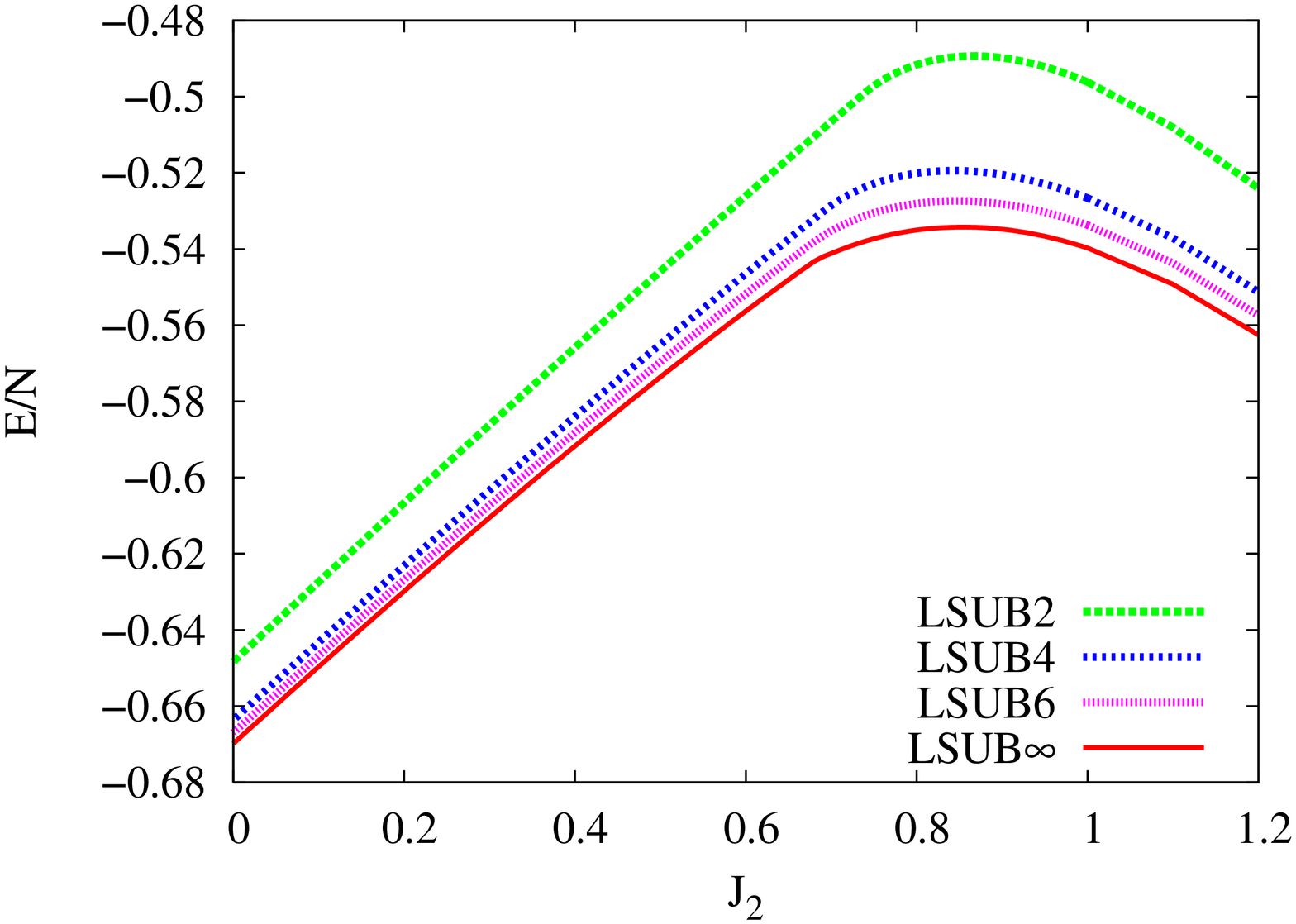,angle=270}}}
 }
\mbox{
 \subfloat[$n=\{3,5,7\}$]{\scalebox{0.3}{\epsfig{file=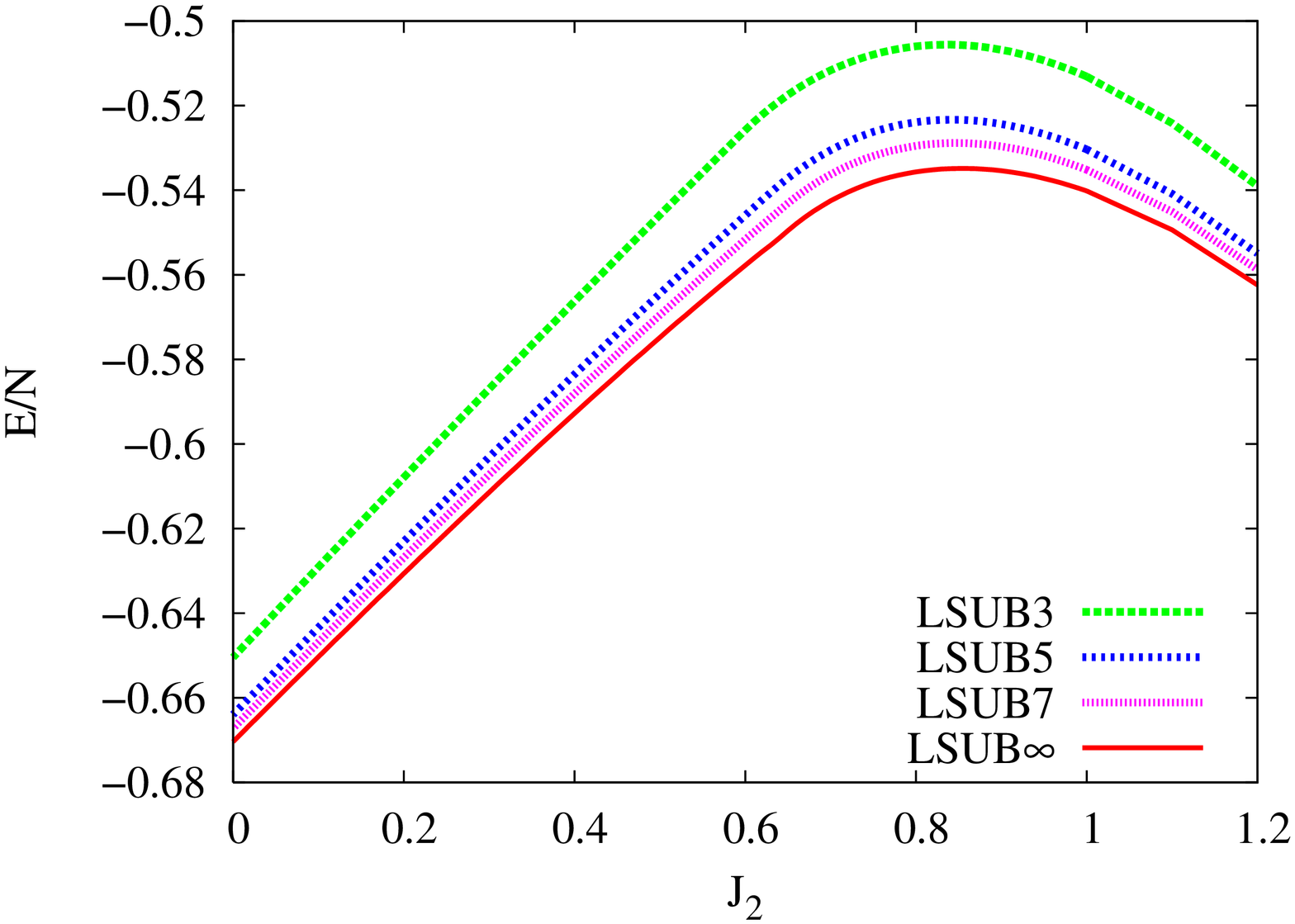,angle=270}}}
}
\caption{(Color online) Ground-state energy per spin versus $J_{2}$
  for the N\'{e}el and canted phases of the spin-1/2 Union Jack
  Hamiltonian of Eq.\ (\ref{H}) with $J_{1} \equiv 1$. The CCM results using
  the canted model state are shown for various LSUB$n$ approximations with (a)
  ($n=\{2,4,6\}$ and (b) ($n=\{3,5,7\}$) with the canting angle
  $\phi=\phi_{{\rm LSUB}n}$ that minimizes $E_{{\rm LSUB}n}(\phi)$.
  We also show the $n \rightarrow \infty$ extrapolated result from
  using Eq.\ (\ref{Extrapo_E}).}
\label{E}
\end{figure*}

In Fig.~\ref{E}
we show the CCM results for the gs energy per spin in
various LSUB$n$ approximations based on the canted (and N\'{e}el)
model states, with the canting angle $\phi_{{\rm LSUB}n}$ chosen to
minimize the energy $E_{{\rm LSUB}n}(\phi)$, as shown in
Fig.~\ref{angleVSj2}. We also show separately the extrapolated
(LSUB$\infty$) results obtained from Eq.~(\ref{Extrapo_E}) using the
separate data sets $n=\{2,4,6\}$ and $n=\{3,5,7\}$ as shown. As is
expected from our previous discussion the energy curves themselves
show very little evidence of the phase transition at $\kappa =
\kappa_{c_{1}}$, with the energy and its first derivative seemingly
continuous.

\begin{figure*}[t]
\mbox{
   \subfloat[$n=\{2,4,6\}$\label{M_even}]{\scalebox{0.3}{\epsfig{file=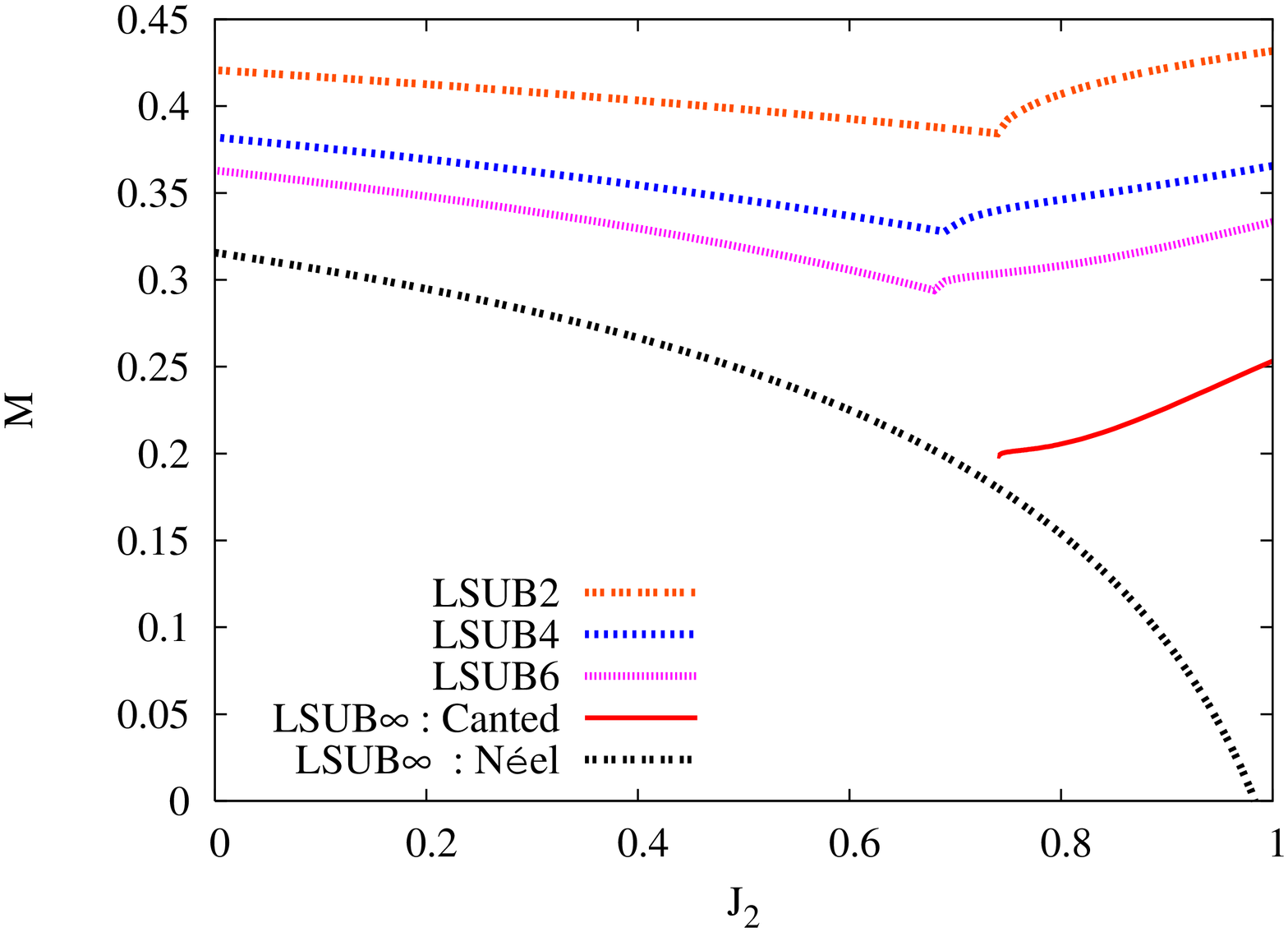,angle=270}}}
}
\mbox{
   \subfloat[$n=\{3,5,7\}$\label{M_odd}]{\scalebox{0.3}{\epsfig{file=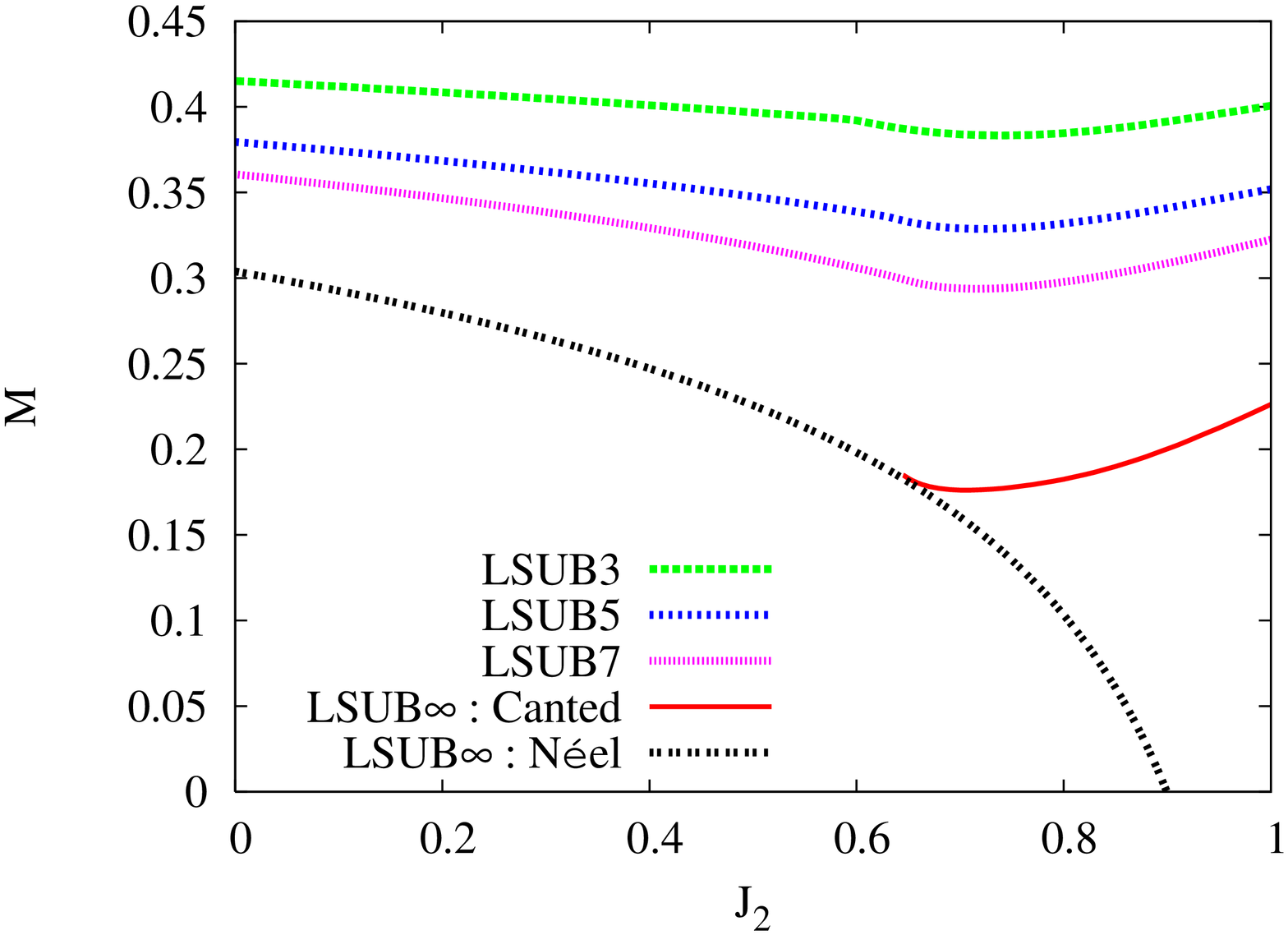,angle=270}}} 
}
\caption{(Color online) Ground-state magnetic order parameter
  (i.e.,~the average on-site magnetization) versus $J_{2}$ for the
  N\'{e}el and canted phases of the spin-1/2 Union Jack Hamiltonian of
  Eq.\ (\ref{H}) with $J_{1} \equiv 1$. The CCM results using the
  canted model state are shown for various LSUB$n$ approximations with
  (a) ($n=\{2,4,6\}$ and (b) $n=\{3,5,7\}$) with the canting angle
  $\phi=\phi_{{\rm LSUB}n}$ that minimizes $E_{{\rm LSUB}n}(\phi)$. We
  also show the $n \rightarrow \infty$ extrapolated result from using
  Eq.\ (\ref{Extrapo_M}).}
\label{M}
\end{figure*}

Much clearer evidence for the transition between the N\'{e}el and
canted phases is observed in our corresponding results for the gs
magnetic order parameter $M$ (the average on-site magnetization) shown
in Fig.~\ref{M}.
For the raw LSUB$n$ data we display the results for the N\'{e}el phase
only for values of $\kappa < \kappa$$^{{\rm LSUB}n}_{c_{1}}$ for
clarity. However, the extrapolated (LSUB$\infty$) results for the
N\'{e}el phase are shown for all values of $\kappa$ in the range shown
and where $M > 0$, using the extrapolation scheme of
Eq.~(\ref{Extrapo_M}) and the LSUB$n$ results based on the N\'{e}el
model state. Once again we extrapolate the odd-$n$ and even-$n$
LSUB$n$ results separately. It is interesting to note from
Fig.~\ref{M} that the raw LSUB$n$ data show the transition at
$\kappa^{{\rm LSUB}n}_{c_{1}}$ more clearly for even values of $n$
than for odd values of $n$. For the canted phase (for which
$\phi_{{\rm LSUB}n} \neq 0$) we can clearly only show the extrapolated
(LSUB$\infty$) results using Eq.~(\ref{Extrapo_M}), for regions of
$\kappa$ for which we have data for all of the set $n=\{2,4,6\}$ or
$n=\{3,5,7\}$. We see from Table~\ref{table_CritPt} that for the
even-$n$ values we are limited (by the LSUB2 results) to values
$\kappa > \kappa^{{\rm LSUB}2}_{c_{1}} \approx 0.740$, whereas for the
odd-$n$ values we are limited (by the LSUB7 results) to values $\kappa
> \kappa^{{\rm LSUB}7}_{c_{1}} \approx 0.645$. The separate odd-$n$
LSUB$\infty$ extrapolation curves for $M$ for the N\'{e}el and canted
phases are seen from Fig.~\ref{M_odd} to be extremely close at the
value $\kappa = 0.645$ and the curves appear to be about to meet at an
angle which is either zero or very close to zero. A straightforward
LSUB$\infty$ extrapolation using Eq.~(\ref{Extrapo_M}) for the three
whole LSUB$n$ curves (N\'{e}el plus canted) with $n=\{3,5,7\}$ shows a
value $\kappa_{c_{1}} \approx 0.637$ at which the extrapolated curve
diverges (at zero or very small angle) from the corresponding
LSUB$\infty$ estimate for the N\'{e}el state shown in
Fig.~\ref{M_odd}. For the corresponding even-$n$ data shown in
Fig.~\ref{M_even}, simple extrapolations of the LSUB$\infty$ curve to
lower values of $\kappa < \kappa^{{\rm LSUB}2}_{c_{1}} \approx 0.740$
using simple cubic or higher-order polynomial fits in $\kappa$ give a
corresponding estimate of $\kappa_{c_{1}} \approx 0.680$ at which the
N\'{e}el and canted phases meet. Both the even-$n$ and odd-$n$ LSUB$n$
extrapolations yield a nonzero value for the average on-site
magnetization of $M \approx 0.195 \pm 0.005$ at the phase transition
point $\kappa_{c_{1}}$. Thus the evidence from the behavior of the
order parameter is that the transition at $\kappa_{c_{1}}$ is a
first-order one, in the sense that the order parameter does not go to
zero at $\kappa_{c_{1}}$, although it is certainly continuous at this
point, and with every indication that its derivative as a function of
$\kappa$ is also continuous (or very nearly so) at $\kappa =
\kappa_{c_{1}}$.

\begin{figure*}[t]
\mbox{
   \subfloat[$n=\{2,4,6\}$]{\scalebox{0.3}{\epsfig{file=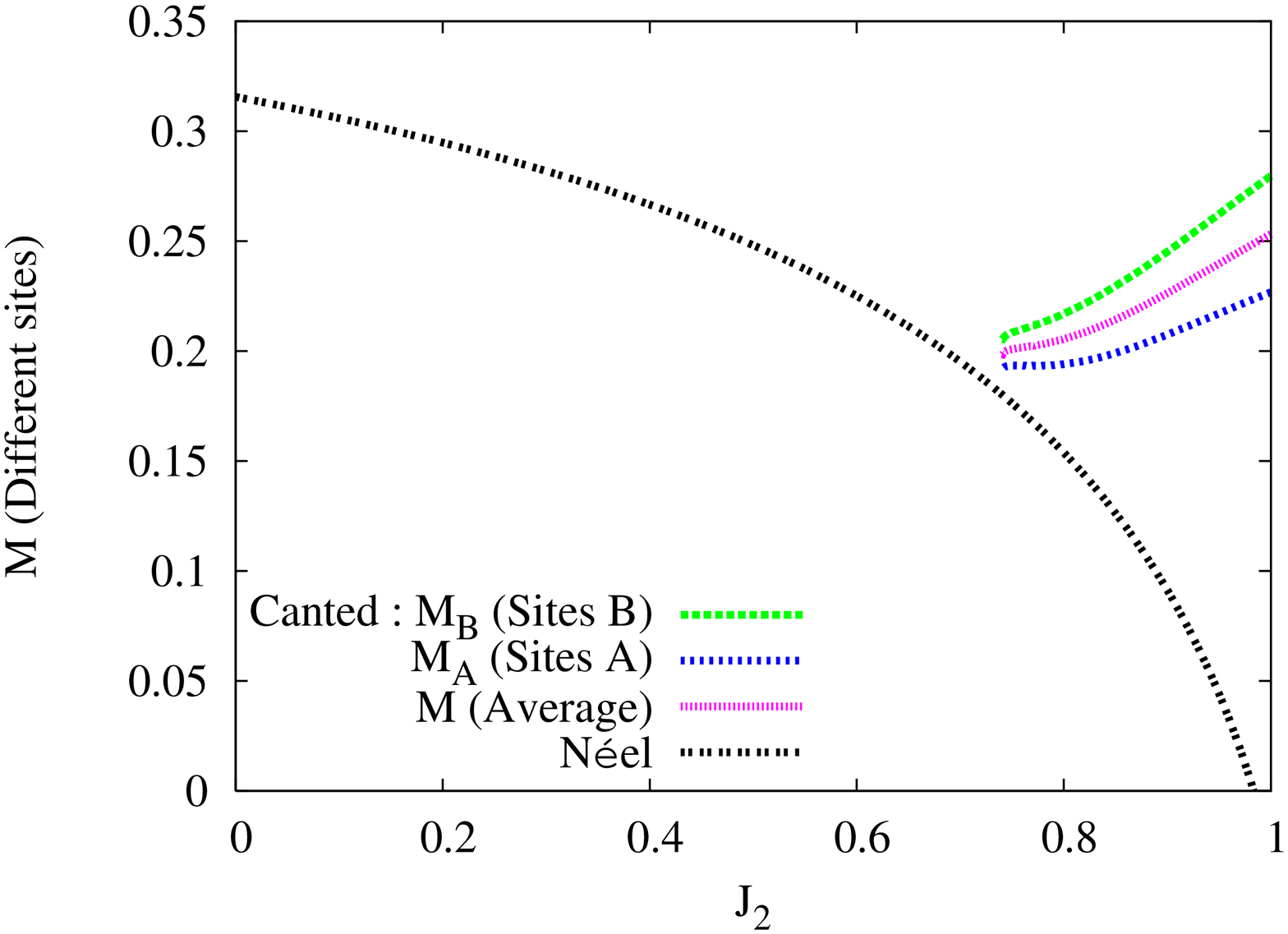,angle=270}}}
}
\mbox{
   \subfloat[$n=\{3,5,7\}$]{\scalebox{0.3}{\epsfig{file=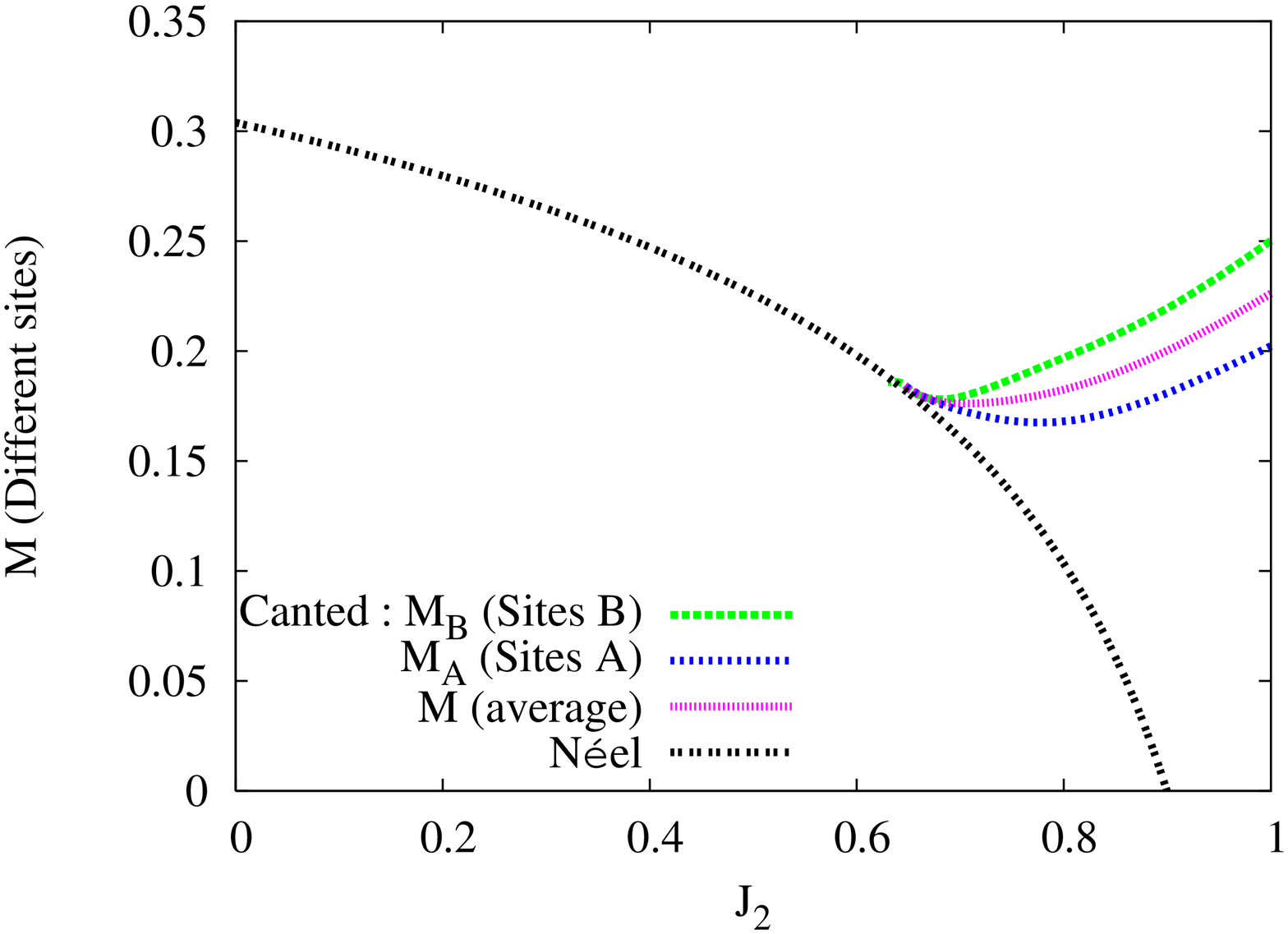,angle=270}}} 
}
\caption{(Color online) Extrapolated curves (LSUB$\infty$) for the
  ground-state magnetic order parameters (i.e., the on-site
  magnetizations) $M_{{\rm A}}$ at sites A (joined by 8 bonds to other
  sites) and $M_{{\rm B}}$ at sites B (joined by 4 bonds to other sites) of the Union Jack
  lattice [and see Fig.~\ref{canted_model}] versus $J_{2}$ for the N\'{e}el
  and canted phases of the spin-1/2 Union Jack Hamiltonian of Eq.\
  (\ref{H}) with $J_{1} \equiv 1$. The CCM results using the canted model
  state are shown for various LSUB$n$ approximations ($n=\{2,4,6\}$
  and $n=\{3,5,7\}$) with the canting angle $\phi=\phi_{{\rm LSUB}n}$
  that minimizes $E_{{\rm LSUB}n}(\phi)$.}
\label{M_DiffSites}
\end{figure*}

We also show in Fig.~\ref{M_DiffSites}
the corresponding extrapolated (LSUB$\infty$) results for the average
on-site magnetization as a function of $J_{2}$ (with $J_{1} \equiv
1$), or hence equivalently as a function of $\kappa$, for both the A
sites ($M_{{\rm A}}$) and the B sites ($M_{{\rm B}}$) of the Union Jack lattice. We recall that, as
shown in Fig.~\ref{canted_model}, each of the A and B sites is
connected to 4 NN sites on the square lattice by $J_{1}$-bonds,
whereas each of the A sites is additionally connected to 4 NNN sites
on the square lattice by $J_{2}$-bonds. The extrapolations are shown
in exactly the same regions, and for the same reasons, as those shown
in Fig.~\ref{M}.

\begin{figure*}[t]
\epsfig{file=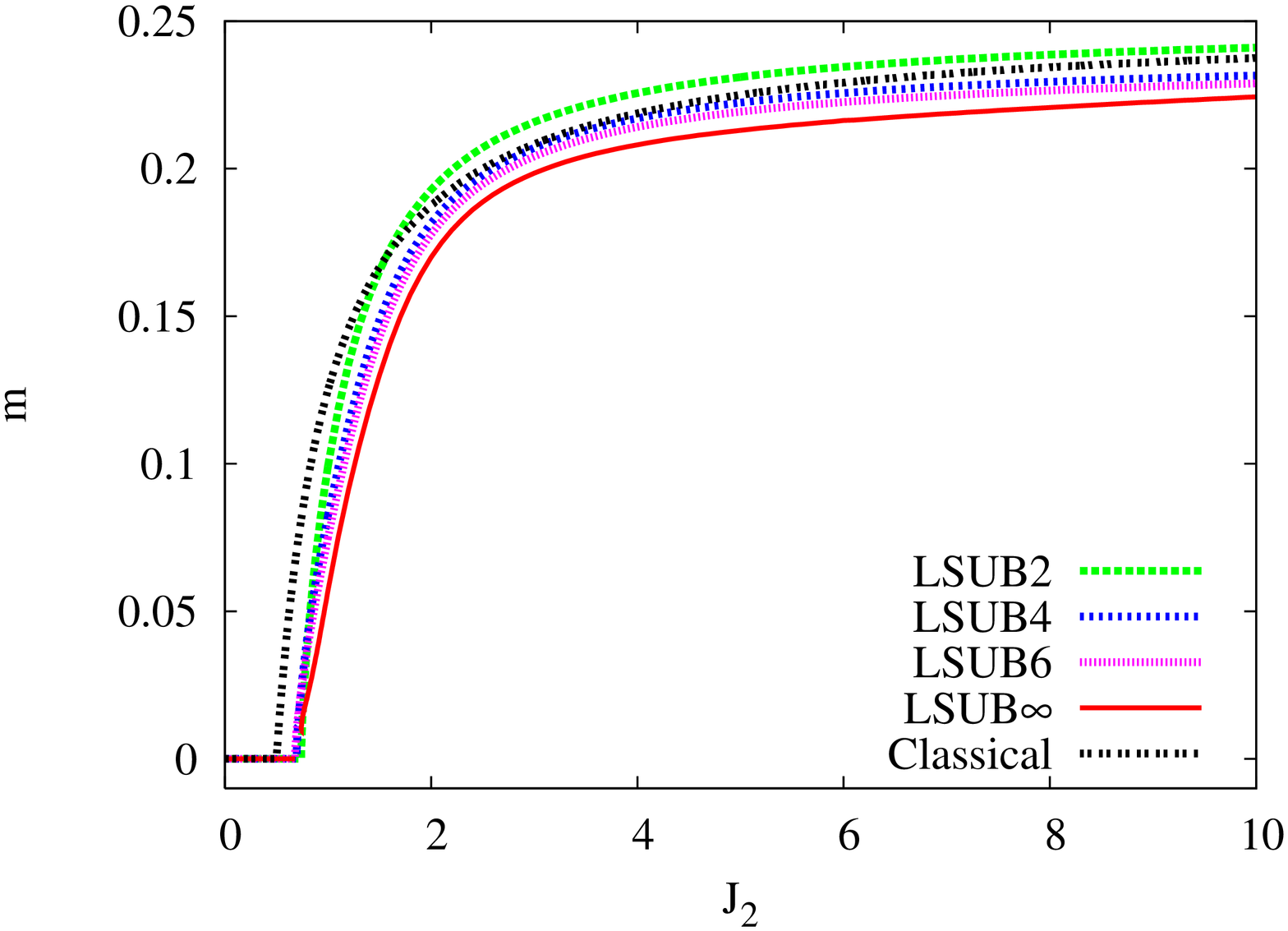,width=6cm,angle=270}
\caption{(Color online) The
  total ground-state magnetization per site, $m=M_{{\rm B}}-M_{{\rm A}}\,{\rm cos}\,\phi$, of the Union Jack
  lattice versus $J_{2}$ of the spin-1/2 Union Jack Hamiltonian of Eq.\
  (\ref{H}) with $J_{1} \equiv 1$. The CCM results using the canted model
  state are shown for various LSUB$n$ approximations ($n=\{2,4,6\}$) with the canting angle $\phi=\phi_{{\rm LSUB}n}$
  that minimizes $E_{{\rm LSUB}n}(\phi)$. We also show the $n \rightarrow \infty$ extrapolated result from using Eq.~(\ref{Extrapo_M}) and compare it with the classical value $m^{{\rm cl}}=\frac{1}{4}(1-{\rm cos}\,\phi_{{\rm cl}})$.}
\label{totalM}
\end{figure*}

In Fig.~\ref{totalM} 
we also show the total gs magnetization per site, $m=\frac{1}{2}(M_{{\rm
    B}}-M_{{A}}\,{\rm cos}\,\phi)$, from using our CCM LSUB$n$ results with
$=\{2,4,6\}$ and with the canting angle $\phi = \phi_{{\rm LSUB}n}$
that minimizes $E_{{\rm LSUB}n}\,(\phi)$. Clearly $m=0$ in the N\'{e}el
phase where $M_{{\rm A}}=M_{{\rm B}}$ and $\phi=0$. We also show in
Fig.~\ref{totalM} the corresponding classical result $m^{{\rm
    cl}}=\frac{1}{2}s(1-{\rm cos}\,\phi_{{\rm cl}}) =
\frac{1}{4}[1-(2\kappa)^{-1}]$ in the canted phase (with
$s=\frac{1}{2}$). A comparison of the extrapolated (LSUB$\infty$) CCM
curve with its classical counterpart shows very clearly that the
quantum fluctutions for this spin-half Union Jack model modify the
classical behavior only relatively modestly, providing further evidence
to what we also noted earlier in relation to Fig.~\ref{angleVSj2}.

We also comment briefly on the large-$J_{2}$ behaviour of our results
for the canted phase. (We note that for computational purposes it is
easier to re-scale the original Hamiltonian of Eq.~(\ref{H}) by
putting $J_{2} \equiv 1$ and considering small values of $J_{1}$.) The
most interesting feature of the CCM results using the canted state as
model state is that in all LSUB$n$ approximations with $n > 2$ a
termination point $\kappa^{{\rm LSUB}n}_{t}$ is reached, beyond which
no real solution can be found, very similar to the termination points
shown in Fig.~\ref{EvsAngle}. For the even-$n$ sequence the values are
$\kappa^{{\rm LSUB}4}_{t} \approx 80$ and $\kappa^{{\rm LSUB6}}_{t}
\approx 80$, whereas for the odd-$n$ sequence the values are
$\kappa^{{\rm LSUB}3}_{t} \approx 250$, $\kappa^{{\rm LSUB}5}_{t}
\approx 85$, and $\kappa^{{\rm LSUB}7}_{t} \approx 55$.
This is a first indication that the canted state becomes unstable at
very large values of $\kappa$ against the formation of another (as yet
unknown) state, as we discuss further in
Sec.~\ref{Canted_vs_semiStripe} below.

Extrapolations of the gs energy using the data before the terminations
points $\kappa^{{\rm LSUB}n}_{t}$ show that at large $J_{2}$ values we
have $E/N \rightarrow -0.3349J_{2}$ using the even-$n$ LSUB$n$ series
and $E/N \rightarrow -0.3352J_{2}$ using the odd-$n$ series. These
numerical coefficients are precisely half of the values quoted in
Table~\ref{table_EandM_results} for the case $J_{2}=0$. This is
exactly as expected since both the $\kappa \rightarrow 0$ and the
$\kappa \rightarrow \infty$ limits of the Union Jack model are the
square-lattice HAF, where in the latter case the square lattice
contains only half the original sites, namely the A sites. Similarly,
the extrapolated LSUB$\infty$ values at larger values of $\kappa$
before the termination point for the on-site magnetization on the A
sites are $M_{{\rm A}} \rightarrow 0.317$ for the even-$n$ LSUB$n$
series and $M_{{\rm A}} \rightarrow 0.306$ for the odd-$n$
series. Both values are again remarkably consistent with those shown
in Table~\ref{table_EandM_results} for the $J_{2} = 0$ limit. The
corresponding asymptotic values for the B-site magnetization are
consistent with $M_{{\rm B}} \rightarrow 0.5$, as expected for large
values of $J_{2}$.

\subsection{Canted state versus the semi-striped state}
\label{Canted_vs_semiStripe}
We turn finally to our CCM results based on the use of the semi-striped
state shown in Fig.~\ref{SemiStripe_model} as the model state. Unlike
in the case of the corresponding use of the canted state as model
state, the results based on the semi-striped state do not 
terminate at a high value of $J_{2}$ (with $J_{1} \equiv 1$). We found
no indication of such a termination value at any LSUB$n$ level of
approximation for $2 \leq n \leq 7$ for values of $J_{2} < 1000$. All
indications are thus that the semi-striped state is stable out to the
$J_{2} \rightarrow \infty$ limit. Indeed for these LSUB$n$ levels the
CCM solutions based on the semi-striped state as model state exist for
all values $J_{2} > 1$. For example, the LSUB6 solution based on the
semi-striped state terminates at a lower end-point $J_{2} \approx 0.41$.

\begin{figure}
\epsfig{file=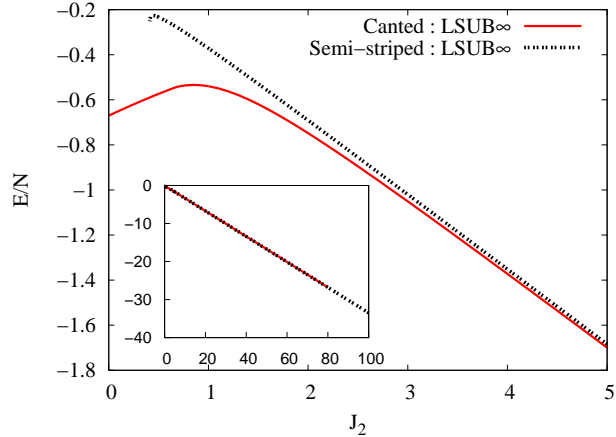,width=6cm,angle=270}
\caption{(Color online) Comparison of the extrapolated (LSUB$\infty$)
  curves for the ground-state energy per spin of the spin-1/2 Union
  Jack model, using the LSUB$n$ data with $n=\{2,4,6\}$ fitted to
  Eq.~(\ref{Extrapo_E}), for the CCM based on the canted and semi-striped
  states as model states. The LSUB4 and LSUB6 approximations (and
  hence also the extrapolated curve) for the canted model state
  terminate at $J_{2} \approx 80$.}
\label{E_Canted_vs_semiSripe}
\end{figure}

In Fig.~\ref{E_Canted_vs_semiSripe}
we compare the extrapolated (LSUB$\infty$) values of the gs energy per
spin, based in each case on the LSUB$n$ results with $n=\{2,4,6\}$,
for our CCM results using the canted and semi-striped states. Although
results for the canted model state become unavailable for $J_{2}
\gtrsim 80$ for the LSUB4 and LSUB6 approximations, the results based
on the canted state lie lower in energy than those based on the
semi-striped state for all values of $J_{2} \lesssim 80$ for which both
sets of solutions exist. Although this is disappointing at first
sight, the two sets of curves become extremely close for larger values
of $J_{2}$ as can be seen from
Fig.~\ref{E_Canted_vs_semiSripe}. Furthermore, we have also attempted
a simple power-law extrapolation of the quantity $E/(NJ_{2}$) for the
gs energy of the canted state in powers of $1/J_{2}$, beyond the
large-$J_{2}$ LSUB$n$ termination points (viz. at $J_{2} \approx 80$
for the LSUB6 approximation). Fits to sixth-, seventh-, and
eighth-order polynomials give virtually identical results for values
of $J_{2}$ in the range $80 \lesssim J_{2} \lesssim 500$ and these
extrapolated curves do indicate that there is a second phase
transition at $\kappa_{c_{2}} \approx 125 \pm 5$ between the canted
and semi-striped phases, such that for values $\kappa > \kappa_{c_{2}}$
the semi-striped phase becomes lower in energy.

The gs energy in both the canted and semi-striped phases approaches the
asymptotic value $E/N \approx -0.3349J_{2}$ for large values of
$J_{2}$ as $J_{2} \rightarrow \infty$ (with $J_{1} \equiv 1$). The
corresponding asymptotic ($J_{2} \rightarrow \infty$) values for the average
on-site magnetization of the semi-striped state are $M_{{\rm A}}
\rightarrow 0.317$ for the A sites and $M_{{\rm B}} \rightarrow 0.5$
for the B sites.

\section{Discussion and Conclusions}
\label{discussion}
In this paper we have used the CCM to study the influence of quantum
fluctuations on the zero-temperature gs phase diagram of a frustrated
spin-half Heisenberg antiferromagnet (HAF) defined on the 2D Union
Jack lattice. We have studied the case where the NN $J_{1}$ bonds are
antiferromagnetic ($J_{1} > 0$) and the competing NNN $J_{2} \equiv
\kappa J_{1}$ bonds in the Union Jack array have a strength in the
range $0 \leq \kappa < \infty$. On the underlying bipartite
square lattice there are thus two types of sites, viz. the A sites
that are connected to the 4 NN sites on the B-sublattice with $J_{1}$-
bonds and to the 4 NNN on the A-sublattice with $J_{2}$-bonds, and the
B sites that are connected only to the 4 NN sites on the A-sublattice
with $J_{1}$-bonds. The $\kappa = 0$ limit of the model thus
corresponds to the spin-half HAF on the original square lattice (of A
and B sites), while the $\kappa \rightarrow \infty$ limit corresponds to the
spin-half HAF on the square lattice comprised of only A sites. We have
seen that at the classical level (corresponding to the case where the
spin quantum number $s \rightarrow \infty$) this Union Jack model has
only two stable gs phases, one with N\'{e}el order for $\kappa <
\kappa$$^{{\rm cl}}_{c} = 0.5$ and another with canted ferrimagnetic order
for $\kappa > \kappa$$^{{\rm cl}}_{c}$. We have therefore first used these
two classical states as CCM model states to investigate the effects of
quantum fluctuations on them.

For the spin-half model we find that the phase transition between the
N\'{e}el antiferromagnetic phase and the canted ferrimagnetic phase
occurs at the higher value $\kappa_{c_{1}} = 0.66 \pm 0.02$. The
evidence from our calculations is that the transition at
$\kappa_{c_{1}}$ is a subtle one. From the energies of the two phases
it appears that the transition is either second-order, as in the
classical case, or possibly, weakly first-order. However, on neither
side of the transition at $\kappa_{c_{1}}$ does the order parameter
$M$ (i.e., the average on-site magnetization) go to zero. Instead as
$\kappa \rightarrow \kappa_{c_{1}}$ from either side, $M \rightarrow
0.195 \pm 0.005$, which is more indicative of a first-order
transition. Furthermore, the slope $dM/d\kappa$ of the average on-site
magnetization as a function of $\kappa$ also seems to be either
continuous or to have only a very weak discontinuity at $\kappa =
\kappa_{c_{1}}$.

Before continuing with the possibility of a further phase we compare
our results with those from previous calculations of the same model
using spin-wave theory (SWT)\cite{Co:2006_PRB,Co:2006_JPhys} and the
linked-cluster series expansion (SE) method.\cite{Zh:2007}  Collins {\it
  et al}.\cite{Co:2006_PRB,Co:2006_JPhys} used linear (or
leading-order) spin-wave theory (LSWT) to show that on the basis of a
comparison of the gs energies of the two phases, the phase transition
between the N\'{e}el and canted phases is of first-order type and
occurs at $\kappa_{c_{1}} \approx 0.84$. In LSWT the N\'{e}el
staggered magnetization per site $M$ remains substantial at this estimate of
$\kappa_{c_{1}} \approx 0.84$. However, it is well known that LSWT
results become unreliable near the transition region, and they
surmised that the N\'{e}el order parameter $M$ might vanish at or
before this point, yielding a possible scenario where a second-order
N\'{e}el transition might occur at a value $\kappa_{c_{1}} \lesssim
0.84$, followed by a possible intermediate spin-liquid phase (as in
the pure $J_{1}$--$J_{2}$ model, as discussed in Sec.~\ref{Introd}),
and then a first-order transition to the canted phase at a somewhat
larger value of $\kappa$. Our own results provide no evidence at all
for such an intermediate spin-liquid phase between the N\'{e}el
antiferromagnetic and canted ferrimagnetic phases.

Although LSWT is known to give a reasonable description of the
spin-1/2 Heisenberg antiferromagnet on the square lattice
($\kappa=0$), it is surely unable to model the frustrated or
intermediate regime accurately. Similar shortcomings of spin-wave
theory (SWT) have been noted by Igarashi\cite{Ig:1993} in the context
of the related spin-1/2 $J_{1}$--$J_{2}$ model on the square lattice,
discussed briefly in Sec.\ \ref{Introd}. He showed that whereas its
lowest-order version (LSWT) works well when $J_{2}=0$, it consistently
overestimates the quantum fluctuations as the frustration
$J_{2}/J_{1}$ increases. In particular he showed, by going to higher
orders in SWT in powers of $1/s$ where $s$ is the spin quantum number
and LSWT is the leading order, that the expansion converges reasonably
well for $J_{2}/J_{1} \lesssim 0.35$, but for larger values of
$J_{2}/J_{1}$, including the point $J_{2}/J_{1}=0.5$ of maximum
classical frustration, the series loses stability. He also showed that
the higher-order corrections to LSWT for $J_{2}/J_{1} \lesssim 0.4$
make the N\'{e}el-ordered phase more stable than predicted by LSWT. He
concluded that any predictions from SWT for the spin-1/2
$J_{1}$--$J_{2}$ model on the square lattice are likely to be
unreliable for values $J_{2}/J_{1} \gtrsim 0.4$. It is likely that a
similar analysis of the SWT results for the spin-1/2 Union Jack model
studied here would reveal similar shortcomings of LSWT as the
frustration parameter $\kappa \equiv J_{2}/J_{1}$ is increased.

In a later paper\cite{Zh:2007} by Zheng {\it et al}., SE techniques
were applied to our spin-half Union Jack model and were compared with
those from both LSWT for both the N\'{e}el and canted phases and
modified second-order SWT for the N\'{e}el phase. Using the SE method
for the N\'{e}el phase gave what these authors termed very clear
evidence of a second-order phase transition at a critical coupling
$\kappa_{c_{1}} = 0.65 \pm 0.01$ at which the N\'{e}el staggered
magnetization per site vanished. For higher couplings the system was
seen to lie in the canted phase, with no sign of any intermediate
spin-liquid phase between these two magnetically ordered states. Use
of the SE method in the canted phase produced a gs energy which
continues smoothly from the N\'{e}el into the canted phase. Zheng {\it
  el al}.\cite{Zh:2007} found, furthermore, that in the canted phase
the staggered magnetizations per site in both the vertical and horizontal
directions shown in Fig.~\ref{canted_model} also appear to drop
smoothly toward zero around the same value $\kappa_{c_{1}} = 0.65 \pm
0.01$, albeit with very large error bars.

The above SE estimate for $\kappa_{c_{1}}$ is clearly in excellent
agreement with our own. However, whereas the evidence from the order
parameter $M$ from the SE technique clearly favors a second-order
transition at $\kappa_{c_{1}}$ at which $M \rightarrow 0$ from both
sides, our own CCM calculations clearly favor a first-order transition
at which $M \rightarrow 0.195 \pm 0.005$. We note, however, that the
errors on the SE estimates for $M$ become increasingly large as the
phase transition at $\kappa_{c_{1}}$ is approached from either
side. We believe that this could easily account for the seeming
discrepancy between our respective predictions for the order of the
phase transition at $\kappa_{c_{1}}$. We note too that Zheng {\it el
  al.}\cite{Zh:2007} were themselves puzzled by the discrepancy
between the prediction of SWT that the N\'{e}el magnetization per site
$M$ does not vanish at $\kappa_{c_{1}}$ and that of the SE technique
that $M$ vanish does vanish there. While they recognized (as do we, as
we discussed above) that SWT cannot be taken as an infallible quide,
they found the huge difference with the prediction from the SE
technique perturbing. Those authors ended by stating that, in their
opinion, the nature of the transition from the N\'{e}el to the canted
phase in the spin-half Union Jack model deserved further
exploration. We believe that our own work reported here has
considerably illuminated the transition at $\kappa_{c_{1}}$.

Neither SWT nor SE techniques have been applied to the possible
semi-striped state of Fig.~\ref{SemiStripe_model} for the spin-half
Union Jack model and so we have no results against which to compare
our own. We were led to consider such a state as a possible gs phase of
the model at large values of $\kappa$ as discussed in
Sec.~\ref{model_section}. Thus, to recapitulate, the $\kappa
\rightarrow \infty$ limit of the canted phase of the Union Jack model
(for either the quantum $s=1/2$ model considered here or the
classical $s \rightarrow \infty$ case) gives a state in which the
spins on the antiferromagnetically-ordered A-sublattice are orientated at
90$^{\circ}$ to those on the ferromagnetically-ordered
B-sublattice. The actual $\kappa \rightarrow \infty$ limit should, in either
case, be decoupled antiferromagnetic (A) and ferromagnetic (B)
sublattices, with complete degeneracy at the classical level for all
angles of relative ordering directions between the two sublattices. We
argued that quantum fluctuations could, in principle, lift this
degeneracy by the well-known order by disorder
phenomenon.\cite{Vi:1977} Since quantum fluctuations are also well
known from many spin-lattice problems to favor collinearity, there is
a strong {\it a priori} possibility that the semi-striped state of
Fig.~\ref{SemiStripe_model} might be energetically favored at large
values of $\kappa$ over the non-collinear state which is the $\kappa
\rightarrow \infty$ limit of the canted state in which $\phi
\rightarrow 90^{\circ}$.

Accordingly we repeated our CCM calculations using the semi-striped
state as model state. We found some evidence that at very large values
of $\kappa$ there might indeed be a second phase transition at
$\kappa_{c_{2}} \approx 125 \pm 5$, based on the relative energies of
the canted and semi-striped states. Such a prediction is based,
however, on an extrapolation of the data on the canted state into
regimes where the CCM equations have no solution for LSUB$n$
approximations with $n > 3$, and hence cannot be regarded as being as
reliable as our prediction for $\kappa_{c_{1}}$. If the phase
transition at $\kappa_{c_{2}}$ does exist it would be of first-order
type according to our results. It would be of considerable interest to
explore the possible transition at $\kappa_{c_{2}}$ between the canted
and semi-striped phases by other techniques, possibly including SWT
and SE methods.

As has been noted elsewhere,\cite{Bi:2008_JPCM} high-order CCM results
of the sort presented here 
have been seen to provide accurate and
reliable results for a wide range of
such highly frustrated spin-lattice models. Many previous
applications of the CCM to unfrustrated spin models have given
excellent quantitative agreement with other numerical methods
(including exact diagonalization (ED) of small lattices, quantum Monte
Carlo (QMC), and series expansion (SE) techniques). A typical example is
the spin-half HAF on the square lattice, which is the $\kappa=0$ limit
of the present model (and see Table~\ref{table_EandM_results}). It is
interesting to compare for this $\kappa=0$ case, where comparison can
be made with QMC results, the present CCM extrapolations of the
LSUB$n$ data for the infinite lattice to the $n \rightarrow \infty$
limit and the corresponding QMC or ED extrapolations for the results
obtained for finite lattices containing $N$ spins that have to be
carried out to give the $N \rightarrow \infty$ limit. Thus, for the
spin-1/2 HAF on the square lattice the ``distance'' between the CCM
results for the ground-state energy per spin\cite{Fa:2008} at the
LSUB6 (LSUB7) level and the extrapolated LSUB$\infty$ value is
approximately the same as the distance of the corresponding QMC
result\cite{Ru:1992} for a lattice of size $N=12 \times 12$ ($N=16
\times 16$) from its $N \rightarrow \infty$ limit. The corresponding
comparison for the magnetic order parameter $M$ is even more striking.
Thus even the CCM LSUB6 result for $M$ is closer to the LSUB$\infty$
limit than any of the QMC results for $M$ for lattices of $N$ spins
are to their $N \rightarrow \infty$ limit for all lattices up to size
$N=16 \times 16$, the largest for which calculations were
undertaken.\cite{Ru:1992} Such comparisons show, for example, that
even though the ``distance'' between our LSUB$n$ data points for $M$
and the extrapolated ($n \rightarrow \infty$) LSUB$\infty$ result
shown in Fig.\ \ref{M} may, at first sight, appear to be large, they
are completely comparable to or smaller than those in alternative
methods (where those other methods can be applied). Furthermore, where such
alternative methods can be applied, as for the spin-1/2 HAF on the
square lattice, the CCM results are in complete agreement with them.

By contrast, for frustrated spin-lattice models in two dimensions both
the QMC and ED techniques face formidable difficulties. These arise
in the former case due to the ``minus-sign problem'' present for
frustrated systems when the nodal structure of the gs wave function is
unknown, and in the latter case due to the practical restriction to
relatively small lattices imposed by computational limits. The latter
problem is exacerbated for incommensurate phases, and is compounded
due to the large (and essentially uncontrolled) variation of the
results with respect to the different possible shapes of clusters of a
given size.

For highly frustrated spin-lattice models like the present Union Jack
model, a powerful numerical method, complementary to the CCM, is the
linked-cluster series expansion (SE)
technique.\cite{j1j2_square_series1,j1j2_square_series2,j1j2_square_series3,j1j2_square_series4,j1j2_square_series5,We:1999,Pa:2008,He:1990,Ge:1990,Ge:1996}
The SE technique has also been applied to the present
model.\cite{Zh:2007} Our own results have shed considerable light on
the nature of the phase transition at $\kappa_{c_{1}}$ observed by SE
techniques and the discrepancies between the results from SE and SWT
methods.

We end by remarking that it would also be of interest to repeat the
present study for the case of the $s>1/2$ Union Jack model.  The
calculations for this case are more demanding due to an increase at a
given LSUB$n$ level of approximation in the number of fundamental
configurations retained in the CCM correlation operators.
Nevertheless, we hope to be able to report results for this system in
the future.

\section*{ACKNOWLEDGMENTS}
We thank the University of Minnesota Supercomputing Institute for
Digital Simulation and Advanced Computation for the grant of
supercomputing facilities, on which we relied heavily for the
numerical calculations reported here. We are also grateful for the use
of the high-performance computer service machine (i.e., the Horace
clusters) of the Research Computing Services of the University of
Machester.


\begin{thebibliography}{99}




\bibitem{2D_magnetism_1} 
{\em Quantum Magnetism}, 
edited by U.~Schollw{\"{o}}ck, J.~Richter, D.J.J.~Farnell, and R.F.~Bishop, 
Lecture Notes in Physics {\bf 645} (Springer-Verlag, Berlin, 2004).

\bibitem{2D_magnetism_2} 
G.~Misguich and C.~Lhuillier, 
in {\it Frustrated Spin Systems}, edited by H.~T.~Diep (World Scientific, Singapore, 2005), p.~229.

\bibitem{2D_magnetism_3} 
J.B.~Parkinson and D.J.J.~Farnell
in {\it An Introduction to Quantum Spin Systems}, Chapter 11
Lecture Notes in Physics (Springer-Verlag, Berlin, 2010) -- in press.



\bibitem{Sa:1997}
A.~W.~Sandvik,
Phys.\ Rev.\ B {\bf 56}, 11678 (1997).
                                             

\bibitem{square_swt} C.J. Hamer, Z. Weihong, and P. Arndt,  Phys. Rev. B {\bf 46}, 6276 (1992).


\bibitem{square_series} 
Z.~Weihong, J. Oitmaa, and C. J. Hamer, Phys. Rev. B {\bf 43}, 8321 (1991).


\bibitem{square_manousakis} E. Manousakis, Rev. Mod. Phys. {\bf 63},  1 (1991).


\bibitem{j1j2_square_ccm1} R.F. Bishop, D.J.J. Farnell, and J.B. Parkinson, Phys. Rev. B {\bf 58},  6394 (1998).


                                
\bibitem{Bi:2008_PRB} 
R.~F.~Bishop, P.~H.~Y.~Li, R.~Darradi, J.~Schulenburg, and J.~Richter,
Phys.\ Rev.\ B {\bf 78}, 054412 (2008).


\bibitem{Bi:2008_JPCM} 
R.~F.~Bishop, P.~H.~Y.~Li, R.~Darradi, and J.~Richter, 
J.\ Phys.: Condens.\ Matter {\bf 20}, 255251 (2008).



\bibitem{j1j2_square_ccm4} R. Darradi, J. Richter, J. Schulenburg,
  R.~F.~Bishop, and P.~H.~Y.~Li, J.~Phys.: Conf. Ser. {\bf 145}, 012049
  (2009).

\bibitem{j1j2_square_ccm5} R. Darradi, O. Derzhko, R. Zinke, J. Schulenburg, S. E. Kr\"uger, and J. Richter, Phys. Rev. B {\bf 78}, 214415 (2008).

\bibitem{j1j2_square_series1} J. Oitmaa and Z.~Weihong, Phys. Rev. B {\bf 54}, 3022 (1996).

\bibitem{j1j2_square_series2} R.R.P. Singh, Z.~Weihong, C. J. Hamer, and J. Oitmaa, Phys. Rev. B {\bf 60}, 7278 (1999).

\bibitem{j1j2_square_series3} V. N. Kotov, J. Oitmaa, O. Sushkov, and Z. Weihong, Philos.
Mag. B {\bf 80}, 1483 (2000).

\bibitem{j1j2_square_series4}  J. Sirker, Z. Weihong, O. P. Sushkov, and J. Oitmaa, Phys. Rev.
B {\bf 73}, 184420 (2006).

\bibitem{j1j2_square_series5}  T. Pardini and R.R.P. Singh, Phys. Rev. B {\bf 79}, 094413 (2009).

\bibitem{j1j2_square_ed1} E. Dagotto and A. Moreo, Phys. Rev. Lett. {\bf 63}, 2148 (1989); Phys. Rev. B {\bf 39} 4744(R) (1989).

\bibitem{j1j2_square_ed2} H.J. Schulz and T.A.L. Ziman, Europhys. Lett. {\bf 18}, 355 (1992); H.J. Schulz, T.A.L. Ziman,
and D. Poilblanc J. Phys. I {\bf 6}, 675 (1996).

\bibitem{j1j2_square_ed3} J. Richter and J. Schulenburg, arXiv:0909.3723v1.

\bibitem{j1j2_square_mf} L. Isaev, G. Ortiz, and J. Dukelsky, Phys. Rev. B {\bf 79}, 024409 (2009).







\bibitem{shastry1} B. S. Shastry and  B. Sutherland, Physica B  {\bf 108}, 1069 (1981).


\bibitem{shastry2} R. Darradi, J. Richter, and D.J.J. Farnell, Phys. Rev. B. {\bf 72}, 104425 (2005).

\bibitem{shastry3} D.J.J. Farnell, J. Richter, R. Zinke, and R.F. Bishop, J.~Stat.~Phys.~{\bf 135}, 175 (2009) 

\bibitem{square_triangle}
R.~F.~Bishop, P.~H.~Y.~Li, D.~J.~J.~Farnell, and C.~E.~Campbell,
Phys.\ Rev.\ B {\bf 79}, 174405 (2009).


\bibitem{Co:2006_PRB}
A.~Collins, J.~McEvoy, D.~Robinson, C.~J.~Hamer, and Z.~Weihong, 
Phys.\ Rev.\ B {\bf 73}, 024407 (2006);
{\it ibid.} {\bf 75}, 189902(E) (2007).

\bibitem{Co:2006_JPhys}
A.~Collins, J.~McEvoy, D.~Robinson, C.~J.~Hamer, and Z.~Weihong, 
J.\ Phys.: Conf. Ser. {\bf 42}, 71 (2006).


\bibitem{Zh:2007}
W.~Zheng, J.~Oitmaa, and C.~J.~Hamer,
Phys.\ Rev.\ B {\bf 75}, 184418 (2007);
{\it ibid}. {\bf 76}, 189903(E) (2007).

















\bibitem{Vi:1977} 
J.~Villain, 
J.\ Phys.\ (France) {\bf 38}, 385 (1977); 
J.~Villain, R.~Bidaux, J.~P.~Carton, and R.~Conte, 
{\it ibid.} {\bf 41}, 1263 (1980).










\bibitem{Bi:1991} 
R.~F.~Bishop, 
Theor.\ Chim.\ Acta {\bf 80}, 95 (1991).

\bibitem{Bi:1998} 
R.~F.~Bishop,  
in {\em Microscopic Quantum Many-Body Theories and Their Applications}, 
edited by J.~Navarro and A.~Polls, {\em Lecture Notes in Physics} {\bf 510} 
(Springer-Verlag, Berlin, 1998), p.1.

\bibitem{Fa:2004} 
D.~J.~J.~Farnell and R.~F.~Bishop, 
in {\em Quantum Magnetism}, 
edited by U.~Schollw{\"{o}}ck, J.~Richter, D.~J.~J.~Farnell, and R.~F.~Bishop, 
{\em Lecture Notes in Physics} {\bf 645} (Springer-Verlag, Berlin, 2004), p.307.


\bibitem{Fa:2001}
D.~J.~J.~Farnell, R.~F.~Bishop, and K.~A.~Gernoth,
Phys.\ Rev.\ B {\bf 63}, 220402(R) (2001).

\bibitem{Kr:2000} 
S.~E.~Kr{\"{u}}ger, J.~Richter, J.~Schulenburg, D.~J.~J.~Farnell, and R.~F.~Bishop, 
Phys.\ Rev.\ B {\bf 61}, 14607 (2000).

\bibitem{Schm:2006} 
D.~Schmalfu$\ss$, R.~Darradi, J.~Richter, J.~Schulenburg, and D.~Ihle, 
Phys.\ Rev.\ Lett. {\bf 97}, 157201 (2006).



\bibitem{Ze:1998} 
C.~Zeng, D.~J.~J.~Farnell, and R.~F.~Bishop, 
J.\ Stat.\ Phys. {\bf 90}, 327 (1998).



\bibitem{ccm} 
We use the program package ``Crystallographic Coupled
  Cluster Method'' (CCCM) of D.~J.~J.~Farnell and J.~Schulenburg, see
  http://www-e.uni-magdeburg.de/jschulen/ccm/index.html.

\bibitem{Fa:2008} 
D.~J.~J.~Farnell and R.~F.~Bishop,
Int.\ J.\ Mod.\ Phys.\ B {\bf 22}, 3369 (2008).

\bibitem{Mo:1953}
P.~M.~Morse and H.~Feshbach,
{\it Methods of Theoretical Physics}, Part II (McGraw-Hill, New York, 1953).

\bibitem{Ko:1996}
H.~Kontani, M.~E.~Zhitomirsky, and K.~Ueda,
J.\ Phys.\ Soc.\ Jpn.\ {\bf 65}, 1566 (1996).




\bibitem{Ma:1955}
W.~Marshall, 
Proc.\ R.\ Soc.\ London, Ser.\ A {\bf 232}, 48 (1955).


\bibitem{Ig:1993}
J.~Igarashi,
J.\ Phys.\ Soc.\ Jpn.\ {\bf 62}, 4449 (1993).

\bibitem{Ru:1992}
K.~J.~Runge,
Phys.\ Rev.\ B {\bf 45}, 7229 (1992); 12292 (1992).           

\bibitem{He:1990}
H.~X.~He, C.~J.~Hamer, and J.~Oitmaa,
J.\ Phys.\ A {\bf 23}, 1775 (1990).

\bibitem{Ge:1990}
M.~P.~Gelfand, R.~R.~P.~Singh, and D.~A.~Huse, 
J.\ Stat.\ Phys. {\bf 59}, 1093 (1990).

\bibitem{Ge:1996}
M.~P.~Gelfand, 
Solid State Commun.\ {\bf 98}, 11 (1996).

\bibitem{We:1999}
Z.~Weihong, R.~H.~McKenzie, and R.~R.~P.~Singh,
Phys.\ Rev.\ B {\bf 59}, 14367 (1999).

\bibitem{Pa:2008}
T.~Pardini and R.~R.~P.~Singh,
Phys.\ Rev.\ B {\bf 77}, 214433 (2008).





\end{thebibliography}
\end{document}